\documentstyle[12pt]{article}
\def\9{\phantom 0}      
\renewcommand\linebreak{\unskip\break} 
\def\Re{{\cal R \mskip-4mu \lower.1ex \hbox{\it e}\,}}
\def\Im{{\cal I \mskip-5mu \lower.1ex \hbox{\it m}\,}}
\def\ie{{\it i.e.}}
\def\eg{{\it e.g.}}

\def\etal{{\it et al.}}
\def\ibid{{\it ibid}.}
\def\sub#1{_{\lower.25ex\hbox{$\scriptstyle#1$}}}
\def\sul#1{_{\kern-.1em#1}}
\def\sll#1{_{\kern-.2em#1}}
\def\sbl#1{_{\kern-.1em\lower.25ex\hbox{$\scriptstyle#1$}}}
\def\ssb#1{_{\lower.25ex\hbox{$\scriptscriptstyle#1$}}}
\def\sbb#1{_{\lower.4ex\hbox{$\scriptstyle#1$}}}

\def\to{\rightarrow}
\def\mh{\ifmmode m\sbl H \else $m\sbl H$\fi}
\def\mch{\ifmmode m_{H^\pm} \else $m_{H^\pm}$\fi}
\def\mt{\ifmmode m_t\else $m_t$\fi}
\def\mc{\ifmmode m_c\else $m_c$\fi}
\def\mz{\ifmmode M_Z\else $M_Z$\fi}
\def\mw{\ifmmode M_W\else $M_W$\fi}
\def\mws{\ifmmode M_W^2 \else $M_W^2$\fi}
\def\mhs{\ifmmode m_H^2 \else $m_H^2$\fi}
\def\mzs{\ifmmode M_Z^2 \else $M_Z^2$\fi}
\def\mts{\ifmmode m_t^2 \else $m_t^2$\fi}
\def\mcs{\ifmmode m_c^2 \else $m_c^2$\fi}
\def\mchs{\ifmmode m_{H^\pm}^2 \else $m_{H^\pm}^2$\fi}
\def\ztwo{\ifmmode Z_2\else $Z_2$\fi}
\def\zone{\ifmmode Z_1\else $Z_1$\fi}
\def\mtwo{\ifmmode M_2\else $M_2$\fi}
\def\mone{\ifmmode M_1\else $M_1$\fi}
\def\tb{\ifmmode \tan\beta \else $\tan\beta$\fi}
\def\xw{\ifmmode x\sub w\else $x\sub w$\fi}
\def\ch{\ifmmode H^\pm \else $H^\pm$\fi}
\def\lum{\ifmmode {\cal L}\else ${\cal L}$\fi}
\def\inpb{\ifmmode {\rm pb}^{-1}\else ${\rm pb}^{-1}$\fi}
\def\infb{\ifmmode {\rm fb}^{-1}\else ${\rm fb}^{-1}$\fi}
\def\epem{\ifmmode e^+e^-\else $e^+e^-$\fi}
\def\ppb{\ifmmode \bar pp\else $\bar pp$\fi}

\def\sss{\scriptscriptstyle}
\def\mx{M_{\sss X}}
\def\smallS{{\sss S}}
\def\smallT{{\sss T}}
\def\smallV{{\sss V}}
\def\ks{k_\smallS}
\def\kv{k_\smallV}

\def\beq{\begin{equation}}
\def\eeq{\end{equation}}

\def\ms{M_{\sss S}}
\def\qs{Q_{\sss S}}
\def\roughly#1{\mathrel{\raise.3ex\hbox{$#1$\kern-.75em\lower1ex\hbox{$\sim$}}}}

\newcommand{\be}{\begin{equation}}
\newcommand{\bea}{\begin{eqnarray}} \newcommand{\eea}{\end{eqnarray}}
\newcommand{\ba}{\begin{array}} \newcommand{\ea}{\end{array}}

\def\bsg{\ifmmode b\rightarrow s\gamma \else $b\rightarrow s\gamma$\fi}

\newskip\zatskip \zatskip=0pt plus0pt minus0pt
\def\matth{\mathsurround=0pt}
\def\lsim{\mathrel{\mathpalette\atversim<}}

\def\atversim#1#2{\lower0.7ex\vbox{\baselineskip\zatskip\lineskip\zatskip
  \lineskiplimit 0pt\ialign{$\matth#1\hfil##\hfil$\crcr#2\crcr\sim\crcr}}}

\def\s{\\ \vspace*{-3mm}}
\def\nn{\noindent}
\newcommand{\non}{\nonumber}
\newcommand{\ee}{e^+ e^-}
\newcommand{\ra}{\rightarrow}

\begin{document}
\def\psfig#1{}
\newlength{\captsize} \let\captsize=\small 
\newlength{\captwidth}                     

\rightline{\vbox{\halign{&#\hfil\cr
&SLAC--PUB--95--6772\cr
&GPP-UdeM-TH-95-17\cr
&TRI--PP-95-05\cr}}}

\vspace*{0.3cm}

\bigskip
\begin{center}
{\Large{\bf NEW PARTICLES AND INTERACTIONS}}

\vskip1.6pc

{\bf
A. Djouadi$^{a}$
J. Ng$^{b}$ and
T. G. Rizzo$^{c}$\footnote{Work supported by the Department of Energy, contract
DE-AC03-76SF00515.}}

\vskip1pc

$^{a}$ {\it Groupe de Physique des Particules, Universit\'e de Montr\'eal, \\
Montr\'eal PQ, H3C 3J7 Canada.} \\
$^{b}$ {\it TRIUMF, Vancouver BC, V6T 2A3 Canada.} \\
$^{c}$ {\it Stanford Linear Accelerator Center, Stanford University,
\\ Stanford, CA 94309 USA.}

\vspace*{0.3cm}

\small

\end{center}

\noindent \underline{Contributors}:
D.\ Atwood,
G.\ Azuelos,
G.\ B\'elanger,
G.\ Bhattacharya,
J.\ Bl\"umlein,
P.\ Chiappetta,
M.\ Doncheski,
R.\ Garisto,
S.\ Godfrey,
C.\ Greub,
H.E.\ Haber,
C.\ Heusch,
J.L.\ Hewett,
R.\ Harris,
K.\ Huiti,
A.\ Kagan,
P.A.\ Kalyniak,
J.L.\ Kneur,
J.\ Layssac,
N.\ Lepor\'e,
D.\ London,
J.\ Maalampi,
P.\ Minkowski,
H.\ Nadeau,
E.\ Nardi,
D.\ Ng,
S.\ Pakvasa,
K.A.\ Peterson,
A.\ Pomarol,
M.\ Raidal,
F.M.\ Renard,
E.\ Roulet,
M.\ Sher,
J.\ Soares,
M.\ Spira,
B.\ Thorndyke,
D.\ Tommasini,
C.\ Verzegnassi.

\vspace*{2cm}

\centerline{\bf ABSTRACT}

\vspace*{0.5cm}

\nn We analyze the manifestations of new matter particles predicted by models
of new physics beyond the Standard Model, at present and future high--energy
colliders.  We  consider both the production of these new particles and some of
their indirect  signatures at $pp$ and $eP$ colliders as well as TeV $\ee$
colliders with their $\ee, e \gamma, \gamma \gamma$ and $e^- e^-$ modes. The
report is arranged into  four main sections plus an overview. These sections
will
deal separately with  exotic and excited fermions, difermions, and new
interactions.

\vspace*{1cm}

\nn To appear as a chapter in {\it Electroweak Symmetry Breaking and
Beyond the Standard Model}, edited by T. Barklow, S. Dawson, H.E. Haber and S.
Siegrist, World Scientific.

\newpage

\section{Overview}

Many theories beyond the Standard Model (SM) of the electroweak and strong
interactions, predict the existence of new matter particles. These new
particles  can be cast into three categories: exotic fermions, excited fermions
and difermions. A fourth category will consist in supersymmetric particles but
these will be discussed in a different report. \s

a) \underline{Exotic Fermions.}  New fermions are predicted by many gauge
extensions of the SM and often, they have the usual lepton and baryon quantum
number but non--canonical SU(2)$_L \times$U(1)$_Y$ quantum numbers, e.g. the
left--handed (LH) components are in  weak isosinglets and/or the right--handed
(RH) components in weak isodoublets. Examples of these exotic fermions are the
following \cite{S0}.

$i)$  Sequential fermions: they exist in the simplest extension of the SM where
one simply has to add to the known fermionic spectrum with its three--fold
replica  a fourth family with the same quantum numbers. The existence of a
fourth  generation is still allowed by experimental data, if the associated
neutrino  is heavy enough \cite{S1}. This heavy neutrino should have a RH
component in  order that one can generate its mass, using the Higgs mechanism,
in a gauge  invariant way.

$ii)$ Vector fermions: these occur for instance in the E$_6$ group \cite{S2},
which is suggested as a low energy limit of superstring theories. In E$_6$,
each fermion generation lies in the representation of dimension {\bf 27}, and
in addition to the fifteen SM chiral fields, twelve new fields are needed to
complete  this representation. Among these, there will be two weak isodoublets
of heavy  leptons, a RH and a LH one.

$iii)$ Mirror fermions: they have chiral properties which are opposite to
those of ordinary fermions, i.e. the RH components are in weak
isodoublets and the LH ones are in weak isosinglets; there is also a
LH heavy neutrino \cite{S3}. These fermions appear in many extensions
of the SM and provide a possible way to restore left--right symmetry at the
scale of the electroweak symmetry breaking; they naturally occur in lattice
gauge theories \cite{S4}.

$i$v) Singlet fermions: these are the most discussed fermions in the
literature,
a prominent example being the SO(10) neutrino \cite{S5,S6A,S6B}. Indeed, in
this
unifying group, which is one of the simplest and most economic extensions of
the
SM, the smallest anomaly--free fermion representation has dimension {\bf 16}.
It
contains a RH neutrino in addition to the 15 Weyl fermions in one fermion
generation; this neutrino is of the Majorana type. Singlet neutrinos, which can
be either of Majorana or Dirac type, and new singlet quarks also occur in
E$_6$ \cite{S2}.

It is conceivable that these fermions, if for instance they are protected by
some symmetry, acquire masses not much larger than the Fermi scale. This is
very likely and even necessary, if the new gauge bosons which are generic
predictions of the unified theories are relatively light \cite{S7}. In the
case of  sequential and mirror fermions (at least in the simplest versions of
the  models where the symmetry and the symmetry breaking pattern is the same as
in the SM), theoretical arguments based on the unitarity of scattering
amplitudes suggest that the masses should not exceed a few hundred GeV
\cite{S8}. These particles could be therefore accessible at the next generation
of colliders \cite{S0,S6A,S6B,SALL,S12}. \s

b) \underline{Excited Fermions.} The existence of excited particles is a
characteristic signal of substructure in the fermionic sector \cite{R0A}.
Indeed, if  the known fermions are composite, they should be regarded as the
ground state  to a rich spectrum of excited states; the latter tumble down via
a
magnetic  type de-excitation to the fundamental particles. In analogy with
systems of  substructure spanning from molecular to atomic then hadronic
classifications,  one hopes to explain in this way the well-ordered pattern of
the fermionic  spectrum with its three-fold replica, although there is not yet
a
satisfactory and predictive dynamical model.

In the simplest phenomenological models, excited fermions are assumed to have
spin and isospin 1/2, and that both their LH and RH components are in weak
isodoublets so that they acquire their masses prior to SU(2)$_L\times$U(1)$_Y$
breaking.  The transition between the excited and the fundamental states
consisting of the ordinary SM particles can be described by an SU(3)$_C\times
$SU(2)$_L  \times$U(1)$_Y$ invariant effective interaction of the magnetic
type.
Hence, the excited particles will have full couplings to the gauge bosons and
therefore can be pair produced at colliders, and  also magnetic--type couplings
to ordinary fermions and gauge bosons (that are inversely proportional to the
compositeness scale $\Lambda$) which will  determine the decay of the excited
states and allow for a new production mechanism: single production in
association with their light partners; see Refs.~\cite{R1,R2}.

The search for excited fermions has been systematically pursued for more than
thirty years without any sucess \cite{R0B}. However, this situation  is
not in conflict with the motivation behind the introduction of excited
particles: compositeness is often invoked as a possible alternative to the SM
description of the electroweak symmetry breaking and it is conceivable that the
first  excitations from the new physics would only be felt at, or above, the
Fermi  scale. Therefore, future colliders operating at such energies will play
an important role in testing this possibility. \s

c) \underline{Difermions.} These are scalar or vector particles (spin 1/2
difermions are also discussed in the context of supersymmetric theories) which
have unusual baryon and/or lepton quantum numbers. Examples of these  particles
are as follows.

$i)$ Leptoquarks (LQ): with B$=\pm 1/3$ and L$=\pm 1$ \cite{R0C}. These
particles are expected in Technicolor models in composite models (where quarks
and leptons are made of the same  subconstituents) as bound states of
quark-lepton pairs and also in Grand Unified models (for instance in the E$_6$
model, the supersymmetric partner of the  exotic colored particle which lies in
the {\bf 27} representation, can have leptoquark quantum numbers).  The
leptoquarks will have the usual gauge couplings to the photon, the $W/Z$
bosons
and gluons (for spin--1 LQ's an anomalous magnetic moment can be  added) and
also Yukawa couplings to lepton--quark pairs which determine their decays. For
not too heavy LQ's, this Yukawa coupling should be chiral in order to  avoid
that leptons acquire a too large magnetic moment \cite{g-2}.

A systematic description of leptoquarks quantum numbers and interactions can be
made, starting from an effective Lagrangian with general dimensionless
SU(3)$_C
\times$SU(2)$_L\times$U(1)$_Y$ invariant couplings and conserved lepton and
baryon numbers \cite{buch}. With fermion number $F=3B+L=0$ (the LQ  couples to
lepton--quark pairs) or 2 (the LQ couples to lepton--antiquark pairs), there
are
10 leptoquarks: 5 scalars and 5 vectors (plus their charge conjugate states),
with electric charges ranging from $1/3$ to $5/3$ in absolute value.  The full
set of these leptoquarks is present in a SU(15) based model of
strong--electroweak unification \cite{Framp1}.

$ii)$ Diquarks: with B$=\pm 2/3$ and L$=0$. They are also predicted in
composite models as bound states of quark pairs, and in Grand Unified models
(in the model based on the E$_6$ symmetry group, the supersymmetric partner of
the  exotic colored particle can also have diquark quantum numbers \cite{S2}).

$iii)$ Dileptons with B$=0$ and L$=\pm2$ \cite{Framp2}. These particles occur
in theories where the electroweak gauge group for leptons is extended from
SU(2)$_L\times$U(1)$_Y$ to SU(3) and baryon and lepton numbers are
conserved. They can appear both as scalar and as vector gauge particles and can
be singly or doubly charged; for instance, doubly charged dilepton gauge bosons
appear  in a SU(15) grand unification model.  Dileptons have couplings to
(ordinary) gauge bosons which are  fixed by gauge invariance, and Yukawa
couplings to leptons which mediate the decays.

All these difermions can have masses not too much larger than the electroweak
symmetry breaking scale and therefore could be accessible at future colliders.

The presence of new physics beyond the Standard Model can manifest itself
not only through the discovery of new particles but also, if the latter
are too heavy to be directly produced, through new interactions which alter the
SM predictions for conventional processes involving the known particles. One
can then have an indirect evidence for new physics at a mass  scale higher
than the one being probed directly. In many cases, these new interactions can
be
expressed as higher dimensional, non-renormalizable, operators written in
terms of the SM fields. In dealing with exotica, there is no limit to what new
interactions may exist so any summary must necessarily be limited in scope.

One of the best known examples of this type is the possibility that the top
quark may have anomalous interactions with the gauge bosons of the SM. Indeed,
due to its large mass, the top quark may play a special role and may be the
first place where non-standard effects will appear. These new interactions for
top naturally divide themselves into those associated with QCD (\ie, modified
$t\bar tg$ and $t\bar tgg$ vertices) and new electroweak couplings with
$W,Z,\gamma$. In the QCD case, assuming $CP$ conservation, the lowest
dimension operator representing new physics is the anomalous chromomagnetic
moment, $\kappa$. A non-zero $\kappa$ at hadron colliders can lead to a
significantly modified top pair production cross section with little effect on
various distributions; the influence of $\kappa$ on single top production is
quite small. At the NLC, this new coupling induces a high energy tail in the
gluon energy distribution for the process $e^+e^- \to t\bar tg$. Both hadron
and $e^+e^-$ colliders can probe these new top quark QCD interactions.
In the electroweak sector, one can look for the effects of a finite charge
radius and magnetic dipole moment for the top quark in both $\gamma\gamma \to
t\bar t$ or $e^-e^+\to t\bar t$ with very high sensitivity.

New four-point interactions between the SM fermions and gauge bosons can occur
in a number of ways in addition to those required by gauge invariance.
The simplest example is a dimension-8 $q\bar q \gamma\gamma$ operator which
can lead to an excess of central diphoton pairs at large invariant mass at
proton colliders. Searches for such interactions can probe compositeness
scales of order several TeV.

The existence of Technicolor-like vector particles that are strongly coupled
to
the SM gauge fields may also make their presence felt at scales below their
direct production thresholds.  Precision measurements at a 1 TeV $\ee$ linear
collider can reveal the effects  of such particles with masses in the few TeV
region.

\vspace*{0.5cm}

In this report we will analyze the manifestations of these new particles and
interactions at future high--energy colliders. We will consider both the
production of these new particles and some of their indirect signatures,
at $pp$ colliders [LHC with $\sqrt{s}=14$ TeV], $eP$ colliders [LEP$\times$LHC
with $\sqrt{s}=1.2$ TeV] and $\ee$ colliders [NLC with $\sqrt{s}=0.5$--1 TeV]
with its $\ee, e \gamma , \gamma \gamma$ and $e^- e^-$ modes. The report is
arranged into four main sections plus the Introduction. These sections will
deal separately with exotic and excited fermions, difermions, and new
interactions.

\newpage

\section{Exotic Fermions}

\subsection{Introduction}

Except for singlet neutrinos which have no electromagnetic and weak
charges, the new fermions couple to the photon and/or to the electroweak gauge
bosons $W/Z$ (and for heavy quarks, to gluons as well) with full strength.
These
couplings allow for the pair production of heavy leptons and quarks; in units
of the proton charge, they are given by ($e^F$ is the electric charge of the
fermion $F$, $I^{F}_{3L}/I_{3R}^F$ the third components of LH/RH isospin and
$s_W^2 =1-c_W^2 \equiv\sin^2 \theta_W$)
\beq
v^F_\gamma = e^F , a^F_\gamma=0 , v^F_Z \equiv v_F = \frac{2I_{3L}^F+2I_{3R
}^F- 4e^F s_W^2}{4s_W c_W} , a^F_Z \equiv a_F = \frac{2I_{3L}^F-2I_{3R}^F}
{4s_W c_W}
\eeq
If they have unconventional quantum numbers, the new fermions  will mix with
the SM fermions which have the same U(1)$_Q$ and SU(3)$_C$ assignments. This
mixing will give rise to new currents which determine to a large extent their
decay properties and allow for a new production mechanism: single production in
association with their light partners. The mixing pattern depends sensitively
on
the considered model and, in general, is rather complicated especially if one
includes the mixing between different generations. However, this
inter--generational mixing should be very small since it would induce at the
tree level, flavor changing neutral currents which are severely constrained by
existing data \cite{S9}.

In the present analysis, we will neglect the inter--generational mixing and
treat the few remaining mixing angles as phenomenological parameters. To
describe our parameterization, let us explicitly write down the interaction of
the electron and its associated neutrino with exotic charged and neutral heavy
leptons. Allowing for both LH and RH mixing, and assuming
small angles so that one can write $\sin\zeta_{L,R} \simeq \zeta_{L,R}$, the
Lagrangian describing the transitions between $e,\nu_e$ and the heavy leptons
$N,E$ of the first generation is ($g_W=e/\sqrt{2}s_W$ and $g_Z= e/2s_Wc_W$)
\begin{eqnarray}
{\cal L} &=& g_W \left[ \zeta_L^{\nu E} \bar{\nu_e}
\gamma_\mu E_L + \zeta_R^{\nu E} \bar{\nu_e} \gamma_\mu E_R \right] W^\mu
+ g_Z \left[ \zeta_L^{eE} \bar{e} \gamma_\mu E_L
+ \zeta_R^{eE} \bar{e} \gamma_\mu E_R \right] Z^\mu \\
&+& g_W \left[ \zeta_L^{eN} \bar{e}
\gamma_\mu N_L + \zeta_R^{eN} \bar{e} \gamma_\mu N_R \right] W^\mu
+ g_Z \left[ \zeta_L^{\nu N} \bar{\nu_e} \gamma_\mu N_L
+ \zeta_R^{\nu N} \bar{\nu}_e \gamma_\mu N_R \right] Z^\mu + h.c. \non
\end{eqnarray}
The generalization to the other lepton families and to quarks is obvious.

Let us now summarize the present experimental constraints on the masses of the
new fermions and on their mixing with the ordinary ones. First, we will  assume
that the new gauge bosons predicted by the Grand Unified Models, will be too
heavy \cite{LU}
to affect the decays and  the production of the exotic fermions. As
previously discussed, we will only allow for a flavor--diagonal mixing; the
latter will alter the couplings of the electroweak gauge bosons to  light
quarks
and leptons from their SM values. Since these couplings  have been very
accurately determined at LEP1 (through the measurement of  total and partial
decay widths as well as forward--backward and  polarization asymmetries) and
in various low-energy experiments   and found to agree with the SM predictions
up to the level of one percent, the  mixing angles are constrained to be
smaller than ${\cal O}(10^{-1})$ \cite{S9}; these constraints are summarized
in Table~1. In the case of leptons, if the LH and RH mixing angles
have the same size, the precise measurement of (g-2)$_{e,\mu}$ leads to even
more stringent constraints, $\zeta <{\cal O}(10^{-2})$ \cite{S10}, so one has
to
set $\zeta_{L,R} \gg  \zeta_{R,L}$.  \s

\def\clel{c_L^{e}}
\def\slel{s_L^{e}}
\def\crel{c_R^{e}}
\def\srel{s_R^{e}}

\def\clmu{c_L^{\mu}}
\def\slmu{s_L^{\mu}}
\def\crmu{c_R^{\mu}}
\def\srmu{s_R^{\mu}}

\def\cltau{c_L^{\tau}}
\def\sltau{s_L^{\tau}}
\def\crtau{c_R^{\tau}}
\def\srtau{s_R^{\tau}}

\def\clu{c_L^{u}}
\def\slu{s_L^{u}}
\def\cru{c_R^{u}}
\def\sru{s_R^{u}}

\def\cld{c_L^{d}}
\def\sld{s_L^{d}}
\def\crd{c_R^{d}}
\def\srd{s_R^{d}}

\def\cls{c_L^{s}}
\def\sls{s_L^{s}}
\def\crs{c_R^{s}}
\def\srs{s_R^{s}}

\def\clc{c_L^{c}}
\def\slc{s_L^{c}}
\def\crc{c_R^{c}}
\def\src{s_R^{c}}

\def\clb{c_L^{b}}
\def\slb{s_L^{b}}
\def\crb{c_R^{b}}
\def\srb{s_R^{b}}

\def\clnue{c_L^{\nu_e}}
\def\slnue{s_L^{\nu_e}}
\def\crnue{c_R^{\nu_e}}
\def\srnue{s_R^{\nu_e}}

\def\clnumu{c_L^{\nu_\mu}}
\def\slnumu{s_L^{\nu_\mu}}
\def\crnumu{c_R^{\nu_\mu}}
\def\srnumu{s_R^{\nu_\mu}}

\def\clnutau{c_L^{\nu_\tau}}
\def\slnutau{s_L^{\nu_\tau}}
\def\crnutau{c_R^{\nu_\tau}}
\def\srnutau{s_R^{\nu_\tau}}

\def\kud{\kappa_{ud}}
\def\kus{\kappa_{us}}
\def\kcd{\kappa_{cd}}
\def\kcs{\kappa_{cs}}

\def\clesq{\left(c_L^e\right)^2}
\def\clnuesq{\left(c_L^{\nu_e}\right)^2}
\def\clmusq{\left(c_L^\mu\right)^2}
\def\clnumusq{\left(c_L^{\nu_\mu}\right)^2}
\def\slesq{\left(s_L^e\right)^2}
\def\slnuesq{\left(s_L^{\nu_e}\right)^2}
\def\slmusq{\left(s_L^\mu\right)^2}
\def\slnumusq{\left(s_L^{\nu_\mu}\right)^2}
\def\sresq{\left(s_R^e\right)^2}
\def\srnuesq{\left(s_R^{\nu_e}\right)^2}
\def\srmusq{\left(s_R^\mu\right)^2}
\def\srnumusq{\left(s_R^{\nu_\mu}\right)^2}
\def\cllisq{\left(c_L^{\ell_i}\right)^2}
\def\clnuisq{\left(c_L^{\nu_i}\right)^2}
\def\sllisq{\left(s_L^{\ell_i}\right)^2}
\def\slnuisq{\left(s_L^{\nu_i}\right)^2}
\def\srlisq{\left(s_R^{\ell_i}\right)^2}
\def\srnuisq{\left(s_R^{\nu_i}\right)^2}
\def\cli{c_L^i}
\def\sri{s_R^i}

\def\slusq{\left(s_L^u\right)^2}
\def\sldsq{\left(s_L^d\right)^2}
\def\slssq{\left(s_L^s\right)^2}
\def\slcsq{\left(s_L^c\right)^2}
\def\clisq{\left(c_L^i\right)^2}
\def\srisq{\left(s_R^i\right)^2}
\def\clfsq{\left(c_L^f\right)^2}
\def\srfsq{\left(s_R^f\right)^2}
\def\sltausq{\left(s_L^\tau\right)^2}
\def\srtausq{\left(s_R^\tau\right)^2}
\def\srusq{\left(s_R^u\right)^2}
\def\srdsq{\left(s_R^d\right)^2}
\def\srssq{\left(s_R^s\right)^2}
\def\srcsq{\left(s_R^c\right)^2}
\def\slbsq{\left(s_L^b\right)^2}
\def\srbsq{\left(s_R^b\right)^2}
\def\slnutausq{\left(s_L^{\nu_\tau}\right)^2}
\def\srqsq{\left(s_R^q\right)^2}

\begin{small}
\nn Table 1: 90\% C.L. upper limits on the ordinary-exotic flavor diagonal
mixing angles for individual fits (one angle at a time is allowed to vary) and
joint fits (all angles allowed to vary simultaneously). Only the results
corresponding to neutrinos mixed with heavy singlet leptons are shown. In most
of the cases, LEP measurements of partial widths and asymmetries give the most
effective constraints. $s_L^e$ corresponds to $\sin\zeta_L^e$, etc..;
$m_t=170\,$GeV and $M_H=200\,$GeV are assumed.
\vspace*{-0.8cm}
\begin{center}
$$
\vbox{\offinterlineskip
\halign
{&\vrule#&
   \strut\quad\hfil#\hfil\quad\cr
\noalign{\hrule}
height2pt&\omit&&\omit&&\omit&&\omit&&\omit&&\omit&\cr
& \omit &&Individual&&Joint&& \omit &&Individual&&Joint&\cr
height2pt&\omit&&\omit&&\omit&&\omit&&\omit&&\omit&\cr
\noalign{\hrule}
height2pt&\omit&&\omit&&\omit&&\omit&&\omit&&\omit&\cr
& $\slesq$ && 0.0016 && 0.0054  &&
 $\slusq$ && 0.0022 && 0.012 &\cr
& $\sresq$ && 0.0020 && 0.0018 &&
 $\srusq$ && 0.010 && 0.023 &\cr
& $\slmusq$ && 0.0013 && 0.0049 &&
 $\sldsq$ && 0.0026 && 0.016 &\cr
& $\srmusq$ && 0.0019 && 0.0040 &&
 $\srdsq$ && 0.0066 && 0.019 &\cr
& $\sltausq$ && 0.0011 && 0.0037 &&
 $\slssq$ && 0.0036 && 0.019 &\cr
& $\srtausq$ && 0.0018 && 0.0034 &&
 $\left(s_R^s\right)^{2}$ && 0.021 && 0.059 &\cr
& \omit && \omit && \omit &&
 $\slcsq$ && 0.0044 && 0.024 &\cr
& $\slnuesq$ && 0.0053 && 0.0053 &&
 $\srcsq$ && 0.0097 && 0.043 &\cr
& $\slnumusq$ && 0.0020 && 0.0052 &&
 $\slbsq$ && 0.0017 && 0.031 &\cr
& $\left(s_L^{\nu_\tau}\right)^{2}$ && 0.0055 && 0.017 &&
 $\left(s_R^b\right)^{2}$ && 0.0091 && 0.015 &\cr
height3pt&\omit&&\omit&&\omit&&\omit&&\omit&&\omit&\cr
\noalign{\hrule}}}
$$
\end{center}
\end{small}

\vspace*{1mm}

{}From the negative search of new states and from the measurement of $Z$ decay
widths at LEP1, one can infer a bound of the order of $M_Z/2$ on the masses of
the new fermions \cite{S1} independently of their mixing, except for singlet
neutrinos. Masses up to $m_F \sim M_W$ can be probed at LEP2.  In the case of
heavy neutrinos, including the gauge singlets, an additional constraint is
provided by the negative search \cite{S11} of these states through single
production in $Z$ decays: if the $\nu N$ mixing angle is of the  order of $\sim
0.1$ or larger, $m_N$ should be larger than $M_W$ \cite{S11};  a similar mass
bound can be established for the heavy  charged lepton. Note that for  mixing
angles much smaller than ${\cal O}(10^{-2})$, no bound can be derived on  the
singlet neutrinos masses: the production cross section is small and/or the
heavy neutrino escapes detection because of its too long decay length.

The heavy fermions decay through mixing into massive gauge bosons plus their
ordinary light partners; for masses larger than $M_W/M_Z$ the vector bosons
will be on--shell. Using the scaled masses $a_V=M_{W,Z}^2/m_F^2$, the
decay widths are \cite{S12}
\begin{eqnarray}
\Gamma (F_{L,R} \ra Vf) &=& \frac{\alpha \delta_V} {32 s_W^2 c_W^2} \left(
\zeta_{L,R}^{fF} \right)^2 \ \frac{m_F^3}{M_Z^2}  ( 1-a_V)^2 (1+2a_V)
\end{eqnarray}
with $\delta_V=1(2)$ for $Z(W)$. For small mixing angles, the heavy fermions
have very narrow widths: for $\zeta_L/\zeta_R \sim 0.1$ and masses around 100
GeV the partial decay widths are less than 10 MeV. The decay widths increase
rapidly for increasing fermion masses, $\Gamma \sim m_F^3$~, but for allowed
values of the mixing angles, they do not exceed the 10 GeV range even for $m_F
\sim {\cal O} (1$~TeV). The charged current decay mode is always dominant and
for fermion masses much larger than $M_Z$, the branching ratios are 1/3 and 2/3
for the neutral and charged current decays, respectively. For Majorana
neutrinos \cite{S6A}, both the $l^-W^+ /\nu_l Z$ and $l^+ W^- / \bar{\nu_l} Z$
are possible; this makes its total decay width twice as large as for Dirac
neutrinos. To fully reconstruct the heavy leptons from their final decay
products one needs the branching ratio of their decays into visible particles,
for large $m_L$ they are given by ($N$ is a Dirac neutrino)
\begin{eqnarray}
Br(E^\pm \ra Z l^\pm \ra jj l^\pm) &\simeq& \ 0.23 \nonumber \\
Br(N \ra W^+ l^- \ra jj l^-) &\simeq & \ 0.43
\end{eqnarray}
In the case of $E$, one can also include the cleaner $Z\ra \ee +\mu^+\mu^-$
decays, but the branching ratio is rather small: $\sim 6\%$ compared to $\sim
70\%$ for $Z \ra$ hadrons. Finally, we note that cascade decays are also
possible: either two leptons are in the same isodoublet and the heaviest one
can
decay into the lighter, or the mixing between heavy leptons is much  larger
that
the heavy--light lepton mixing in which case, the decay of a heavy  lepton into
a lighter one first will be favored; these cascade decays will not be
considered here; for some details see ref.~\cite{S12}.

A possibility that should not be overlooked is that a heavy charged lepton
could
be quite long-lived, with a lifetime long enough to leave a visible track in a
detector \cite{S13} (for long-lived quarks, see $e.g.$ Ref.~\cite{bn}).  A
simple model with such a lepton would be a model in
which the heavy leptons form a vectorlike doublet, and in which mixing with the
lighter generations is either absent or suppressed by ratios of neutrino masses
(which then would give a mixing angle less than $10^{-10}$).  In this case, the
charged lepton and neutrino are degenerate in mass at tree level.  Radiative
corrections will break this degeneracy, and will give a mass splitting of
between 270 and 330 MeV (as the lepton mass ranges from 100 GeV to 800 GeV) ;
this remarkable insensitivity to the lepton mass is reflected in the lepton
lifetime, which ranges from 1.2 to 2.0 nanoseconds. Such a particle would leave
a visible track, then decay into neutrals plus a low energy electron or muon
pair.  With such a clear signature, the discovery reach at a hadron collider
would be much higher than for conventional heavy leptons, and would be at the
kinematic limit for $\ee$ colliders. It turns out that if one assumes that this
doublet lies in a supersymmetric theory, then the resulting lifetime is
virtually independent of the supersymmetry parameter space, and is  unchanged
from the non-supersymmetric case.

Such a model is not particularly unusual \cite{S14}.  For example,
the leptonic
extension of Frampton and Kephart  aspon model (which offers a solution
to the strong CP problem) has such a doublet with
small mixing.  In the model of Griest and Sher, extra generations of Higgs
doublets in SUSY were considered.  Assuming only that
a symmetry suppresses FCNC at tree level, it has been shown that the
additional Higgsinos form a doublet like the above model.  The
extra Higgs bosons themselves have the unusual property that the
second lightest (neutral) is a few hundred MeV heavier than the
lightest (neutral), leading to a low energy electron or muon pair
coming from a point detached from the interaction; here backgrounds
may be considerable, and are under investigation \cite{S13}.

Constraints on a fourth generation SM quarks $(t',b')$ and leptons $(N,E)$
requires a special discussion \cite{S15}. We assume that $m_{t'} > m_t$ and
that there is no Majorana mass for the RH neutrino. First, one needs to
parameterize the CKM matrix for four generations; for a real 4$\times$4 matrix,
one has to introduce three new parameters $V_{u b'} \equiv \epsilon$,
$V_{c b'} \equiv \delta$ and $V_{t b'}\equiv \sin\theta$, and one has
\begin{small}
\begin{equation}
V \simeq \left(
\begin{array}{cccc}
1 & \lambda & A\rho\lambda^3 & \epsilon \\
-\lambda & 1 & A\lambda^2 & \delta\\
A\lambda^3(1-\rho)\cos\theta -(\epsilon -\delta\lambda) \sin\theta &
-A\lambda^2\cos\theta -\delta\sin\theta & \cos\theta & \sin\theta \\
-A\lambda^3(1-\rho)\sin\theta -(\epsilon -\delta\lambda)\cos\theta &
A\lambda^2\sin\theta   -\delta\cos\theta  & -\sin\theta & \cos\theta
\end{array} \right)
\end{equation}
\end{small}
which preserves much of the Wolfenstein parameterization for the 3$\times$3
case.
The present experimental data constrain the Wolfenstein parameters to
$\lambda=0.22, A=0.79 \pm 0.12$ and $|\rho| = 0.36 \pm 0.09$, and the upper
bounds for the new parameters, obtained from unitarity, are given as
$|\epsilon| \leq 0.077$ and $|\delta| \leq 0.594$. From the precision LEP data
(mainly from the $\rho$ parameter and the ratio of the $Z\rightarrow b \bar{b}$
to the hadronic widths), one obtains for $m_t = 150$ GeV and $M_H=300$  GeV,
$m_{b'} \sim m_{t'}$ for $\sin\theta \sim 0$ and $m_{b'}>$ 100 GeV,  $m_{t'} <$
300 GeV for $\sin\theta \sim 1/\sqrt{2}$. The LEP experiment tell us in
addition
that $m_{N}>45$ GeV. CLEO's experimental bound on  $b \rightarrow s \gamma$ as
well as the differences $B_d-{\overline B_d}$ and  $B_s-{\overline B_s}$  would
constrain $\theta$, $m_{t'}$ and also $\delta$  (for CLEO). The branching ratio
for the decay $K^+ \rightarrow \pi^+ \nu \bar{\nu}$  would also place
bounds on $\theta$ and $m_{t'}$ as well as $\epsilon$  and $\delta$;
$D$-$\bar{D}$ mixing would give better constraint on the last  two parameters.
Finally, the heavy-light lepton mixing angles are constrained at the level of
$\sim 0.1$ for the third generation (mainly from lepton  universality) and at a
level of $\sim 10^{-2}$--$10^{-3}$ for the two other generations (from FCNC
processes like $\mu \ra e\gamma , \mu \ra 3e$  and $\mu$-$e$ conversion in
nuclei, etc..).

The fourth generation leptons $N$ and $E$ will decay into leptons and massive
gauge bosons (the decay widths, up to mixing angles, are given in eq.~3).
Because $\sin\theta$ is not
necessary small, the decays of $t',~b'$ as well as $t$ depend on $m_{b'}$. If
$m_{t'} < m_{b'}$, $t'$ will be produced first and will then mainly decay into
$b+W$ or $s+W$ depending on the parameters $\sin\theta$ and $\delta$ (since the
constraints on $\delta$ are not very stringent, $t'\rightarrow s W$ is not
necessary suppressed).  In the opposite case,  $m_{b'} < m_{t'}$, there are two
possibilities: if $m_{b'}>m_t + m_W$, the main decays will be $b' \rightarrow t
W$ and $b' \rightarrow (c,u) W$ (if $\sin\theta > \delta,\epsilon$ the former
mode is dominant, leading to  a nice signature); if $m_{b'}<m_t + m_W$, the
FCNC
decays $b' \rightarrow b  (g,\gamma,Z)$ can be important (if $\delta$ and
$\epsilon$ are small, say less  than $10^{-3}$) compared to the CC decays $b'
\rightarrow (c,u) W$, and $b' \rightarrow b Z \ra b l^+l^-$ would provide a
spectacular signature.

\subsection{Production in $e^+e^-$ Colliders}

If their masses are smaller than the beam energy, the new fermions can be
pair produced in $\ee$ collisions: through $s$--channel $\gamma$ and $Z$
exchange for charged fermions, and only $Z$ exchange for heavy neutrinos.
The differential cross section (with $\theta$ specifying the direction of the
fermion $F$ with respect to the $e^-$) is \cite{S12}
\begin{small}
\begin{equation}
\frac{d \sigma}{d \cos \theta } = \frac{3}{8} \sigma_0N_c \beta_F
\left[ (1+ \cos^2 \theta) (\sigma_{VV} + \beta^2_F \sigma_{AA}) +(1-\beta^2)
\sin^2\theta\sigma_{VV} +2 \beta_F \cos \theta \sigma_{VA}\right] \,,
\end{equation}
\end{small}
\noindent
where $N_{c}$ is a color factor, $\sigma_0=4\pi \alpha^2/3s$ the
point--like QED cross section and $\beta_F= (1-4m_F^2/s)^{1/2}$ is the
velocity of the fermion in the final state; in terms of
the $FF \gamma$ and $FFZ$ couplings and $z=M_Z^2/s$; $\sigma_{VV}$,
$\sigma_{VA}$ and $\sigma_{AA}$  are
\begin{small}
\begin{eqnarray}
\sigma_{VV} =e_e^2 e_F^2 + \frac{2 e_ee_F v_e v_F}{1-z} + \frac{(a_e^2+ v_e^2)
v_F^2}{(1-z)^2} \ , \ \sigma_{AA} = \frac{(a_e^2+ v_e^2)a_F^2}{(1-z)^2} \\ \non
\sigma_{VA} = \frac{2 e_e e_F a_e a_F}{1-z} + \frac{4v_ea_e v_F a_F}
{(1-z)^2} \hspace*{3cm}
\end{eqnarray}
\end{small}
\nn The total production cross sections and forward-backward asymmetries are
then
\begin{small}
\begin{eqnarray}
\sigma_F = \sigma_0 N_c \ \left[ \frac{1}{2} \beta_F (3-\beta_F^2) \sigma_{VV}
+ \beta_F^3 \sigma_{AA} \right] \ \ , \ \
A_F = \frac{3}{4} \beta_F \frac{\sigma_{VA}}{\frac{1}{2}(3-\beta^2_F)
\sigma_{VV}+ \beta^2_F \sigma_{AA}}
\end{eqnarray}
\end{small}
The cross sections for mirror (which are the same for sequential) and vector
isodoublet heavy leptons are
displayed in Fig.~1a for a c.m. energy of 1 TeV; the cross sections for heavy
quarks are of the same order of magnitude as the one for $E$.  As one can see,
they are rather large: with $\int {\cal L}= 100$ fb$^{-1}$ one can expect $\sim
10^{3}-10^{4}$ events. The backgrounds (mainly from three vector boson
production)
have been discussed in \cite{S12}, and are small compared to the signal.
It is therefore clear that the detection of pair--produced heavy
leptons with masses close to the kinematical limit should be straightforward
at $\ee$ colliders. The angular distributions are shown in Fig.~1b, and one
notes that they are symmetric for vector fermions leading to $A_{FB}=0$; for
mirror fermions, $A_{FB}$ is sizeable and has opposite sign compared to
sequential fermions. The polarization 4-vectors of the heavy fermions can
be measured and would also allow to discriminate between mirror, vector and
sequential fermions \cite{S12}.

Charged fermions can also be pair--produced at $\gamma\gamma$ colliders,
the total cross section for unpolarized photon beams is given by
\begin{eqnarray}
\sigma = 3 N_{c}^{f} e_{F}^{4} \sigma_0 \beta_F \ \left[ 1+\beta^2_F +
\frac{1-\beta^4}{2 \beta} \log\frac{1+\beta}{1-\beta} \right]
\end{eqnarray}
It is shown in Fig.~1c for the charged lepton, assuming a fixed c.m energy of
0.8 TeV. Although smaller lepton masses can be probed because of the loss in
energy, the $\gamma \gamma$ mode is interesting since the rates can be much
larger  than in $\ee$ for low $m_F$. For quarks, the charge is
penalizing and the rates are $N_c^f e_Q^4$ times smaller.

\begin{figure}[htbp]
\hspace*{-1cm}
\centerline{\psfig{figure=fig1.ps,height=20.5cm,width=15cm}}
\vspace*{-1.7cm}
\caption{\small Total production cross sections (a) and angular
distributions (b) for the pair production of vector and mirror leptons in
$\ee$ collisions at $\protect\sqrt{s}=$ 1 TeV, and cross section for charged
lepton production in $\gamma \gamma$ collisions at fixed $\protect\sqrt{s}=0.8$
TeV (c).}
\end{figure}

\vspace*{-0.5cm}
\begin{figure}[htbp]
\hspace*{-1cm}
\centerline{\psfig{figure=fig2.ps,height=20.5cm,width=15cm}}
\vspace*{-1.5cm}
\caption{\small Total production cross sections (a), angular distributions
(b) and the longitudinal and transverse components of the polarization vectors
(c,d) for the single production of heavy leptons with LH and RH mixings in
$\ee$ collisions at $\protect \sqrt{s}=$ 1 TeV.}
\end{figure}

In $\ee$ collisions one can also have access to the new fermions via single
production in association with their light partners if the mixing is not too
small \cite{S12}. The process proceeds only via $s$--channel $Z$ exchange in
the
case of  quarks and second/third generation leptons (if inter-generational
mixing is neglected); because $\zeta_{L,R}<0.1$, the cross sections are very
small and  make these states unlikely to be found this way. But for the first
generation  heavy leptons, one has additional $t$--channel exchanges: $W$
exchange for $N$  and $Z$ exchange for $E$, which increase the cross sections
by
several orders  of magnitude. The analytical expressions are quite involved
\cite{S12}, however,  at very high energies, $w=M_W^2/s$ and $z=M_Z^2/s$ can be
neglected and if one  uses $s_W^2 \simeq 1/4$ and assumes $\zeta_{L,R}^{\nu N}
=
\zeta_{L,R}^{e N}= \zeta_{L,R}$, one obtains simple formulae which are good
approximations. Using the scaled mass $a=m_L^2/s$, one has for $\sigma(L)\equiv
\sigma(\ee \ra  L\bar{l})$
\begin{eqnarray}
\sigma(L_{L,R}) = 3 \sigma_0 \left(\zeta_{L,R}\right)^2 (1-a) \kappa (L_{L,R})
\hspace*{4.5cm} \nonumber \\
\kappa(N_L)= \frac{1}{w}+\frac{1}{3} \frac{a-4}{1-a}\log\frac{1-a+w}{w} ,
\kappa(N_R)= \frac{-1}{1+w-a}+\frac{a}{1-a}\log\frac{1-a+w}{w} \hspace*{-3mm}
\nonumber\\
\kappa(E_L)=\kappa(E_R)= \frac{1}{9} \left[ \frac{1-a}{z(1-a+z)}-4\log
\frac{1-a+z}{z} \right] \hspace*{3cm}
\end{eqnarray}
The total cross sections are the same for the charge conjugate states, and for
Majorana neutrinos it is $\sigma= \sigma(L)+\sigma(\bar{L})$ \cite{S6A}. They
are shown  in Fig.~2a for all mixing angles taken to be $\zeta_{L,R}=0.1$. The
cross  sections are very large, especially for $N_L$ where they reach the
picobarn  level. For $E_{L,R}$ they are one order of magnitude smaller, a
consequence of the smaller NC couplings compared to CC couplings.  For $N_R$,
the cross section is  approximately 10 fb across the entire mass range. For
smaller mixing  angles  the rates have to be scaled correspondingly; even for
$E$ and $N_R$, requiring  10 events with $\int  {\cal L}=100$ fb$^{-1}$ for
$m_L=800$ GeV, one can probe $\zeta$ values one order of magnitude smaller.

The angular distributions are shown in Fig.~2b, and one clearly sees that one
can distinguish between neutrinos with LH and RH mixing, and of Dirac or
Majorana type. A further distinction can be made by measuring the final
polarization; for instance the longitudinal and transverse components of the
polarization vectors of $E_L$ and $E_R$ (which cannot be discriminated by the
angular distributions) are practically equal in magnitude but opposite in sign
as shown in Fig.~2c/2d.

To fully reconstruct the heavy lepton masses, the best signals consist of an
$\ee$ pair and two jets for the charged lepton and an $e^\pm$, a pair of jets
and missing momentum for the neutral lepton; the branching ratios are $23\%$
and 43\% respectively. In the case of $E$, the main backgrounds are: $
e^+ e^-\to e^+e^-Z \to \ee jj$, $ e^+ e^- \to Z Z$, $ e^+ e^- \to t
\bar{t} \to W^+W^- jj$ and $ \gamma \gamma \to  e^+ e^-q \bar{q}$. In the
case of $N$, the backgrounds are: $ \ee \to e \nu W$, $ \ee \to WW \to e^\pm
\nu_ jj$ and $ \gamma \gamma \to e (e)  q  \bar{q} $. These backgrounds
can be eliminated or reduced by applying the following cuts for $N(E)$
production: 1) require one and only one $e^\pm$ ($\ee$ pair), 2) the invariant
mass of the two jets should reconstruct to $M_W$ $(M_Z)$, 3) the invariant mass
of the $\ee$ pair should be large (different from $M_Z$), 4) the momentum of
the
neutrino and charged leptons (two charged leptons) should be large, 5) cuts
on the angle between the initial electron and the $W$ $(Z)$ boson, and 6)
a cut $\cos \theta_{l \nu}$ $(\cos \theta_{ll}) <0.5$. Optimizing these
cuts, no events from heavy flavor production or from the $\gamma \gamma$
backgrounds would survive; the backgrounds from vector boson production can be
suppressed to a very low level, while those from single $W/Z$ production
can be a bit higher.

A full simulation \cite{S12,S16} of the signal and backgrounds has been
performed using PYTHIA, for a model detector (an upgraded LEP detector) to
quantify the discovery limits that can be obtained. This simulation was done
assuming a c.m. energy of 500 GeV and an integrated luminosity of 50 fb$^{-1}$.
The signal and background cross sections after applying cuts are shown in
Table 2 (note that at $\sqrt{s}=500$ GeV, the cross sections are practically
the same at 1 TeV, because the dominant contribution comes from the $t$-channel
exchange). The output for heavy leptons with masses of 250, 350 and 450 GeV
and with $\zeta=0.05$ for $E$ and $\zeta_L=0.025$ for $N$ is shown in Fig.~3.
For these $\zeta$ values, one can see that the signal peaks stand out clearly
from the background events, especially for $m_L$ not too close to $\sqrt{s}$.
For $m_E \sim 450$ GeV, only slightly smaller $\zeta$ values can be probed,
while for $m_E \sim 350$ GeV one can go down by at least a factor of two. The
situation is much more favorable for $N_L$, the cross section being one order
of magnitude larger. For $m_L=350$ GeV and requiring that the signal over the
square-root of the background is larger than unity, one can probe mixing
angles down to $\zeta \sim 0.005$ for neutral leptons and $\zeta \sim
0.03$ for charged leptons. At $\sqrt{s}=1$ TeV, these numbers for $m_L$ and
$\zeta^2$ values can be improved by a factor of two.

Finally, we note that heavy fermions cannot be produced singly at $\gamma
\gamma$ colliders (at least in a $2\ra 2$ process); heavy neutral and
charged leptons can be produced in $e\gamma$ collisions in association with
massive gauge bosons \cite{S17}, however only smaller masses can be probed
and the rates are not much larger than in $\ee$ collisions. \s

\nn {\small Table 2: Cross sections for heavy lepton single production and
for the main backgrounds at $\sqrt{s}=0.5$ TeV after successive applications
of cuts; $m_L=350$ GeV and $\zeta= 0.025(0.05)$ are chosen for $N(E)$ and the
masses are in GeV.}

\begin{small}
\begin{table}[hp]
\centering
\begin{tabular}{|c||c|c|c|} \hline
Process & $E^\pm e^\mp$ &  $ e^+ e^- Z $ &   $ Z  Z $  \\ \hline
$ \sigma$ [fb] &    9.5 & 4960 &  615 \\ \hline
$ \times $ B.R. & 2.19 & 3470 & 28.8 \\ \hline
one $\ee$ pair & 1.74 & 93.0 & 23.0 \\
$ 330 < M_E < 370$ & 1.56 & 11.7 & 5.30 \\
$  85 < M_Z < 105$ & 1.41 & 5.84 & 2.87 \\
$ |M_{ll}-M_Z|>12$ & 1.39 & 5.18 & 1.02 \\
$\cos\theta_{ll}<0.5$ & 1.33 & 4.32 & 0.56 \\
$ f( M_E, \cos \theta_{Z})$ & 1.30 & 1.90 &  0.43 \\
 kinem. cuts & 1.30 & 1.55 & 0.39 \\ \hline
\end{tabular}
\hspace*{1mm}
\begin{tabular}{|c||c|c|c|} \hline
Process  &$ \bar{N} \nu $ & $ e \nu W $& $  W  W $  \\ \hline
 $ \sigma$ [fb]                             &    490                     &
8610             &   2600       \\ \hline
 $ \times $ B.R.
                                       &    13.7                    &
5823             &   1140       \\ \hline
  one $e$                              &    13.2                    &
198              &    883       \\
$ 330 < M_N < 370$ &    12.5                    &
11.9             &    100       \\
$  70 < M_W <  90$ &    12.3                    &
10.3             &     70.3     \\
$  M_{l\nu} > 120 $&    11.8                    &
10.0             &     7.93     \\
$ \cos \theta_{l\nu} < 0.5 $              &    11.7                    &
10.0             &     7.80     \\
$ f( M_N, \cos \theta_{Z})$ &    11.7                    &
10.0             &     7.80     \\
 kinem. cuts                           &    11.7                    &
10.0             &     4.13     \\ \hline
 \end{tabular}
\end{table}
\end{small}

\begin{figure}[htbp]
\hspace*{-0.5cm}
\centerline{\psfig{figure=fig3.ps,height=19.5cm,width=15.5cm}}
\vspace*{-0.5cm}
\caption{\small Reconstructed masses for the singly produced heavy leptons
in $\ee \ra \ee j j$ (a) and in $\ee \ra \nu_e e jj$ (b) at $\protect \sqrt{s}
=0.5$ TeV, and for the main backgrounds after the application of the cuts
displayed in Tab.~2.}
\end{figure}

\subsection{New particles in Left--Right and Extended Models}

In this subsection, we will discuss separately the case of the new particles
predicted by Left--Right Models (LRM) based on the symmetry SU(2)$_L
\times$SU(2)$_R \times $U(1)$_{B-L}$ \cite{lrm}. In these models, the LH and
RH fermions
transform as doublets under SU(2)$_L$ and SU(2)$_R$ respectively, and therefore
each generation contains a RH neutrino. The extended symmetry leads also to new
neutral and charged interactions mediated by heavy $Z'$ and $W_R$ gauge bosons.
These new gauge bosons will mix with the SM $Z$ and $W$ bosons, but the mixing
is rather small and can be neglected here. The masses of the new gauge bosons
are constrained to be larger than $\sim 0.5$ TeV; however, in models with
arbitrary Yukawa couplings $M_{W_R}$ can be as low as 300 GeV \cite{LU}.  The
minimal Higgs sector contains doublet and triplet scalar fields which leads to
the existence of neutral, charged and also doubly charged $(\Delta^{++}$) Higgs
bosons. Supersymmetric versions of the LRM have also been considered, and their
particle content is much richer than that of the Minimal Supersymmetric SM. In
particular, there is a doubly charged fermion, the Higgsino $\tilde{\Delta}^
{++}$. We will discuss the manifestation of some of these new particles at
$e^+e^-$ colliders.

The heavy RH neutrinos can be produced in pairs at $e^+e^-$ colliders, $e^+ e^-
\rightarrow NN$, through the $s$--channel exchange of a heavy $Z'$ and
the $t$--channel  exchange of the RH boson $W_{R}$; for Majorana neutrinos, one
has also a $u$--channel $W_R$ exchange. This process has been discussed in
\cite{S6B} where the expressions  for the cross sections can be found. At a 500
GeV $e^+e^-$  collider, they are shown in Fig.~4 for $g_L=g_R$ and
$M_{Z'}$=$M_{W_R}$=1 and 1.5 TeV and for Dirac and Majorana neutrinos. The
cross
section for Dirac is slightly larger than for Majorana neutrinos, especially
close to the kinematical limit due the well--known $\beta^3$ suppression factor
for the latter.

\begin{figure}[htbp]
\centerline{\psfig{figure=fig4.ps,height=7cm,width=12cm,angle=-90}}
\vspace*{0.01cm}
\caption{\small Total cross sections for $\ee \ra NN$ production at a 500 GeV
$\ee$ collider.}
\end{figure}

To probe the Majorana or Dirac nature of the heavy neutrino, one can use the
angular distributions, which are different. $N$ will decay through RH charged
currents into a charged lepton and two--jets. Requiring 20 events
with ${\cal L}=20$ fb$^{-1 }$, one can probe masses of the order of 230 (200)
for Dirac (Majorana) neutrinos for $M_{W_R}=1.5$ TeV. Note that heavy neutrinos
can be also singly produced in $eP$ collisions \cite{S6B} through $W_{R}$
exchange, as will be discussed later. \s

One can look at the situation the other way around, and try to produce  $W$
bosons through the exchange of the heavy neutrinos. An interesting  possibility
is the production of like sign $W$ pairs in $e^-e^-$  collisions{\cite{oldww}}.
This process, if it exists would signal  the existence of new $|\Delta L|=2$
interactions which may manifest themselves  as Majorana masses for neutrinos.
However, it is difficult to generate a  large cross section for this reaction
while simultaneously satisfying the  constraint of tree-level unitarity at
large
values of $s$ and the bounds on  the effective neutrino mass arising from the
lack of observation of  neutrinoless double beta decay. In the LRM \cite{lrm},
as a result of the  see-saw mechanism used to generate small masses for the
ordinary LH  neutrinos, these difficulties can be easily circumvented by
considering the  reaction $e^-e^- \to W_R^-W_R^-$.

The amplitude for $e^-e^- \to W_R^-W_R^-$ gets both $t$- and $u$-channel
contributions from the exchange of the heavy RH neutrino $N$, as well as an
$s-$channel contribution from the exchange of the doubly-charged Higgs boson
$\Delta$ with mass $M_\Delta$ ($W$--$W_R$ mixing is neglected here).
Since the $e^-e^-\Delta$ coupling is proportional to $M_N$ and the $e^-NW_R$
coupling is chiral, the total amplitude is proportional to $M_N$. Thus,
as the Majorana mass of $N$ vanishes so does the amplitude; this is expected
since the $|\Delta L|=2$ interaction is generated by the Majorana mass term.

At energies of $\sqrt{s}=$ 0.5--1.5 TeV, the cross section for  $e^-e^- \to
W_R^-W_R^-$ is quite large, of order a few picobarns, is fairly sensitive to
the values of $M_N$ and $M_\Delta$, and has a rather flat angular distribution.
The $\Delta$ boson may appear as an $s-$channel resonance depending upon the
value of $\sqrt {s}$. Unfortunately, the ``reach" is rather limited since we
are restricted to $M_{W_R}< \sqrt {s}/2$ values, and since
$M_{W_R}$ is heavier than 0.5 TeV and $W_R$ pair production would be not
kinematically accessible at these energies. One has therefore to consider
\cite{tgrww} the possibility of single production of $W_R$ via the reaction
$e^-e^- \to W_R^-(W_R^-)^*\to W_R^-jj$. We limit ourselves to this $jj$
mode to allow for the possibility that $M_N>M_{W_R}$ in which case the
$W_R$ can only decay to jets barring the existence of exotic particles.

Allowing one of the $W_R$'s to be off-shell results in a substantial reduction
in the cross section from the on-shell case to the level of a few fb, which
implies that machine luminosities in the range of ${\cal L}=100-200$ fb$^{-1}$
are required to make use of this channel. The total event rates for the
reaction is shown in Fig.~5, in which we have set $\kappa=g_R/g_L=1$ and scaled
by an
integrated luminosity of 100 fb$^{-1}$. Fig.~5a shows the number of expected
$W_R+jj$ events, as a function of $M_{W_R}$, at a $\sqrt {s}=1$ TeV $e^-e^-$
collider for different choices of $M_N$ and $M_{\Delta}$. The results are seen
to be quite sensitive to the values of these masses.
In Fig.~5b(c), we fix $M_{W_R}=700$ GeV and plot the event rate
as a function of $M_N(M_{\Delta})$ for various values of $M_{\Delta}(M_N)$.
Typically, we see that one can obtain rates of the order of several hundred
events a year, except near the $\Delta$ resonance (where the rates are very
large) or when $M_N$ is small (the cross section vanishes for $M_N=0$).
For most choices of the input masses, one obtains extremely flat angular
distributions, except when $N$ is light in which case a significant angular
dependence is observed as a result of the $t-$ and $u-$ channel poles which
develop as $M_N \ra 0$.
\vspace*{-0.5cm}
\nn
\begin{figure}[htbp]
\centerline{
\psfig{figure=fig5a.ps,height=7.cm,width=8cm,angle=90}
\hspace*{-5mm}
\psfig{figure=fig5b.ps,height=7.cm,width=8cm,angle=90}}
\vspace*{-0.75cm}
\centerline{
\psfig{figure=fig5c.ps,height=7.35cm,width=9.5cm,angle=90}}
\vspace*{-1cm}
\caption{\small Event rates per 100 fb$^{-1}$ for $W_R+jj$ production at a
1 TeV $e^-e^-$ collider a) as a function of $M_R$ for $M_N=M_{\Delta}=1$ TeV
(dots), $M_{\Delta}=1.2$ TeV and $M_N=0.4$ TeV (dashes), $M_{\Delta}=0.3$
and $M_N=0.1$ TeV (dash-dots), $M_{\Delta}=2,~M_N=0.6$ TeV (solid), or
$M_{\Delta}=1.8$ and $M_N=0.6$ TeV (square dots); (b) with $M_R=700$ GeV fixed
as a function of $M_N$ for $M_{\Delta}=0.3(0.6, 1.2, 1.5, 2)$ TeV
corresponding to the dotted(dashed, dash-dotted, solid, square-dotted) curve;
(c) as a function of $M_{\Delta}$ for $M_N=0.2(0.5, 0.8, 1.2, 1.5)$ TeV
corresponding to the dotted(dashed, dash-dotted, solid, square-dotted) curve.}
\end{figure}

\newpage

As mentioned previously, heavy neutrinos may be observed indirectly outside
the
LRM.  One could assume the spontaneous breakdown of e.g. the $E_6$ group via
the chain $E_6 \to SO(10) \to \dots$  down to the SM
group, where the RH $W_R$ bosons (and the doubly charged Higgs bosons) are too
heavy but the two additional  isosinglet neutrinos have masses in the 1--10 TeV
range.  These Majorana neutrinos will mix with light neutrinos, and could yield
the  presently observed mass spectrum {\cite{heusch}}. It has been shown
{\cite{heusch}} that this scenario is open to experimental detection through
the
process $e^-e^- \to W^-W^-$,  where at least the lightest of the two heavy
neutrinos is exchanged.  In contrast to the case where $2W_R$ are produced,
this
process can be observed even at a 0.5 TeV electron--electron collider.

In the kinematic regime well above the $W$ mass, but below that of the heavy
neutrino, where no doubly-charged Higgs contribution is needed for unitarity
reasons (letting us avoid the uncertain couplings of an $L=2$ state to
a flavorless boson), a characteristic energy dependence $\sim s^2$ is shown in
Fig.~\ref{csrange}. The cross-section band displayed there is bounded by the
known limits on heavy-neutrino mixing from rare decay processes and on lepton
universality evidence.

\begin{figure}[htbp]
\let\normalsize=\captsize
\centering{%
\centerline{\psfig{file=heusch.ps,height=6.5cm,width=8cm,clip=}}
}
\caption{%
Cross-section range for the process $e^-e^- \to W^-W^-$ via TeV-mass Majorana
neutrino exchange. The curves indicate extreme values of the neutrino mixing
parameters allowed by present data.
\label{csrange} }
\end{figure}

Note that the spectacular back-to-back $W$ pair decays permit effective
background suppression {\cite{cuypersPL,bargeretal94}}, so that even a moderate
signal may well lead to a convincing discovery; in particular, a change in
incoming electron helicity will eliminate any signal, providing a further test
for its legitimacy.  This is in contrast to the classic discovery channel for
light Majorana neutrinos, neutrinoless double beta decay: here, the heavy
masses
lead to severe signal suppression, and there is no criterion telling a signal
due to light-neutrino exchange from one due to heavy neutrinos {\cite{heusch}}.

Finally, let us consider the Higgs sector and the supersymmetric version of the
LRM \cite{HMRb}. A particularly suitable signal of the  model is offered by the
triplet higgsino  $\tilde\Delta^{++}$, a Dirac fermion consisting of the
fermion
components of  the triplet Higgs fields. As being doubly charged it does not
mix
with other particles, and consequently its mass is given by a single parameter.
The decay modes of the $\tilde\Delta^{++}$ are very limited, since it carries
two units of lepton number and it does not couple to quarks. The
nonconservation of the separate lepton numbers $L_e$, $L_{\mu}$, and $L_{\tau}$
of the $\tilde\Delta^ {++}$ couplings is a very special signature which can be
studied in slepton  pair production where one of the reaction amplitudes
includes $\tilde\Delta^{ --}$ exchange.

The allowed decay modes of the triplet higgsino are $\tilde{\Delta}^{++} \to
\Delta^{++} \lambda^{0}$, $\Delta^{+} \lambda^{+}$,  $\tilde \Delta^{+} W_R^+$,
and $\tilde l^+ l^+$, where $\lambda$'s represent the  fields in the LRM
bi-doublet representation.  In large regions of the parameter space, the
kinematically favoured decay mode is  $\tilde \Delta^{++}  \rightarrow  \tilde
l^+ l^+ $ (this is of course the case only when $m_{\tilde l^+}
<m_{\tilde\Delta^{++}}$, at  least for some slepton flavour, which we will
assume in the following).   If the mass of the triplet Higgs $\Delta$ is of the
order of the SU(2)$_R$ breaking scale,  the first two decay channels are
forbidden energetically in the  case of relatively light triplet higgsinos. For
the same reason  the channel $\tilde \Delta^{+} \, W_R^+$ is kinematically
disfavoured, since the $W_R$  mass is expected to be above 0.5 TeV \cite{LU}.
The decay channel  $\tilde \Delta^{+} W^+$ is supressed by the small $W-W_R$
mixing. In the following we will  assume that $\tilde\Delta^{++}$ and
its charge conjugate state  $\tilde\Delta^{--}$ decay  100\% of the time into
the $\tilde ll$ final states. The charged sleptons $\tilde l$ can decay either
to a charged lepton of the same flavour plus a neutralino, to a neutrino plus a
chargino, or to a charged gauge boson plus a sneutrino:
\beq
\tilde l^+\to l^+ + \tilde\chi_i^0 \ , \ \ \tilde l^+\to \nu + \tilde\chi_i^+
\ , \ \  \tilde l^+\to W^+ + \tilde\nu
\eeq
\noindent
Which  of the various decay channels is the dominant one depends on the mass of
the decaying slepton. The triplet higgsinos can be produced in the next
generation linear electron colliders in the $e^+e^-$, $e^-e^-$, $e^-\gamma$ and
$\gamma\gamma$  collision modes:
\beq
e^+e^-\to \tilde\Delta^{++}\tilde\Delta^{--}, \
e^-e^-\to \tilde\chi^0\tilde\Delta^{--}, \
\gamma e^-\to \tilde\l^+\tilde\Delta^{--}, \
\gamma\gamma \to \tilde\Delta^{++}\tilde\Delta^{--}
\eeq
All these reactions  have a clean experimental signature: a few hard leptons
and missing energy, the background from other processes are thus rather small.
The largest cross section is obtained for the $\gamma \gamma \ra
\tilde\Delta^{++}
\tilde\Delta^{--}$ process; the expression is given in eq.~(9) and because of
the double charge, it is enhanced by a factor 16 compared to the case of a
heavy charged lepton for which it is shown in Fig.~1c.

Finally, the enlarged particle content of the Supersymmetric LRM
also effects other
processes  such as the production of slepton pairs. First, the number of
neutralinos in the $t$-channel production is larger because of the additional
neutral fermions in both the gauge and Higgs sectors. A second difference with
the Minimal Supersymmetric SM case is that there is a new $u$-channel diagram
due to the doubly  charged higgsino discussed above. The combination of these
new contributions  leads to an increase in the production cross section for
slepton pairs in  $e^+e^-$ in comparison to the Minimal Supersymmetric SM
results \cite{HMRb}.

\subsection{Production in $eP$ Collisions}

Heavy leptons of the first generation can be singly produced in $eP$ collisions
\cite{S18,S19} through $t$--channel exchange of a $W$ boson for the neutral and
a $Z$ boson for the charged leptons. One can also have heavy quark production
but the backgrounds are stronger and this process will not be considered here.
For polarized $e$ beams beams, the differential cross section d$\sigma$/d$x$d$y
$ for $e_{L,R}P \ra LX$ $(a=m_L^2/\hat{s})$ is
\begin{small}
\begin{eqnarray}
\frac{d\sigma}{dx dy}= \frac{G_F^2 M_W^4}{2 \pi} \hat{s} \sum_{j,q=L,R}
\zeta_j^2 \left[A_{jq}^L(Q^2) H_{jq}(x,y)q(x,Q^2) + \bar{A}_{jq}^L(Q^2)
\bar{H}_{jq}(x,y)\bar{q}(x,Q^2) \right]
\end{eqnarray}
\vspace*{-5mm}
\begin{eqnarray}
H_{LL}(x,y) &=& \bar{H}_{LR}(x,y) = H_{RR}(x,y) = \bar{H}_{RL}(x,y) = (1-a)
\nonumber \\
H_{LR}(x,y) &=& \bar{H}_{LL}(x,y)=H_{RL}(x,y) = \bar{H}_{RR}(x,y) =
(1-y-a)(1-y)
\nonumber
\end{eqnarray}
\end{small}
Using $a_q=2I^{3L}_{q}$ and $v_q=(2I^{3L}_{q}- 4Q_qs_W^2)$
the expressions of $A^L$ and $\bar{A}^L$  are
\begin{small}
\begin{eqnarray}
A^N_{jq}= \bar{A}^N_{jq} = (Q^2+M_W^2)^{-2} ,
A^E_{Lq}(\bar{A}^E_{Lq})= \bar{A}_{Rq}^E(A_{Rq}^E)=
(Q^2+M_Z^2)^{-2} (v_q \pm a_q)^2/(16c_W^4) \nonumber
\end{eqnarray}
\end{small}
The cross sections are shown in Fig.~7 for $N$ and $E$ production with $\zeta
_{L,R}= 0.1$, at a c.m energy $\sqrt{s}=1.2$ TeV. The largest cross section is
obtained in the  case of $N_L$, closely followed by the one for $N_R$
production; the  cross sections for $E_{L,R}$ are a bit less than one order of
magnitude smaller. The  best signature is $e^-$-jet-jet (see the branching
ratios above); for Majorana  neutrinos, also $e^+W^-$ final states would be
observed \cite{S6A}.  Requiring 20 events  and assuming a luminosity of 2
fb$^{-1}$, one can probe masses up to 700 GeV for $N$ and 550 GeV for $E$,
assuming $\zeta=0.1$.  Note that, together with  the $y$ distributions, the
difference between the longitudinal and transverse components of the
polarizations vectors of the  final leptons can be  used to discriminate
between
particles with LH or RH  mixing \cite{S18}.

For the neutral leptons, the main backgrounds are $W/Z$ production, heavy
flavor
production and higher  order QCD radiation; beam polarization cannot be used to
reduce them. A detailed analysis \cite{S19} has been performed to obtain
allowed
regions  in the $m_N$-$\zeta$ plane. For $\zeta=0.1$ it has been found that the
search  reaches $\sim 800$ GeV at LEP$\times$LHC (with $\sqrt {s}=1.3$ TeV and
${\cal L}=2$ fb$^{-1}$). For HERA ($\sqrt {s}=314$ GeV and ${\cal L}=200$
pb$^{-1}$)  and a HERA upgrade ($\sqrt{s}=450$ GeV and ${\cal L}=4$ fb$^{-1}$),
one reaches  the limits 160 and 320 GeV respectively.

\nn
\vspace*{-3mm}
\begin{figure}[htbp]
\hspace*{-1cm}
\centerline{\psfig{figure=fig6.ps,height=8cm,width=15cm,angle=-90}}
\vspace*{-0.5cm}
\caption{\small Total production cross sections for the single production of
heavy leptons in $eP$ collisions at $\protect \sqrt{s}=$ 1.2 TeV.}
\end{figure}

In the Left--Right symmetric models discussed previously, heavy neutrinos of
the
first generation can be produced in $eP$ collisions via $t$-channel $W_R$
exchange (this  process is somewhat complementary to searches for $W_R$ at the
Tevatron in which $W_R \ra Ne$ and the $N$ decays within the detector). The
reach for this  process was explored in a detailed Monte Carlo study \cite{S19}
under the assumptions  that $g_R=g_L$ and that the CKM matrices in the LH and
RH
sectors are the same. RH beam polarization can be used to reduce backgrounds,
and Majorana decay  signatures for $N$ are required as tags. For HERA, a HERA
upgrade  and LEP$\times$LHC, the discovery reach that has been found
approximately  corresponds to the following  regions in the $m_N-M_{W_R}$
plane:
$m_N+0.34M_{W_R}<276$ GeV, $m_N+0.20M_{W_R} <394$ GeV, and
$m_N+0.38M_{W_R}<1090$ GeV.

\subsection{Production at $pp$ Colliders}

Proton colliders are ideal machines to search for heavy quarks \cite{S20}. The
pair  production subprocesses are gluon--gluon fusion, $gg \ra Q\bar{Q}$ and
quark-antiquark annihilation, $q \bar{q} \ra Q\bar{Q}$, with the gluon fusion
subprocess being by far dominant. The tree--level partonic cross sections  are
well--known \cite{es}
\begin{eqnarray}
\sigma_{gg}(\hat{s}) &=& \frac{\pi \alpha_S^2}{3\hat{s}} \left[ \left( 1+ a
+\frac{a^2}{16} \right) \log \frac{1+\beta}{1-\beta} -\beta \left(\frac{7}{4}
+\frac{31}{16} a \right) \right] \nonumber \\
\sigma_{qq}(\hat{s}) &=& \frac{8\pi \alpha_S^2}{27 \hat{s}} \beta \left(
1+ \frac{1}{2} a \right)
\end{eqnarray}
with $\beta$ the velocity of the quark, $\beta^2=1-a=1-4m_Q^2/\hat{s}$.  The
total cross sections are obtained by integrating over the gluon and quark
luminosities.  One can use the previous tree level formulae in a way so as to
reproduce the full  one--loop corrected $Q \bar Q$ cross-section \cite{nde} in
the limit where the $P_T^j$ of the additional jet tends to zero. The shower
approximation can  be used for this purpose \cite{sa} and the tree-level $Q
\bar
Q + 1$ jet  differential cross-section can be replaced by
\begin{equation}
d\sigma(Q\bar{Q}j)  \longrightarrow d\sigma(Q\bar {Q}j) \times \left[1 -
exp \left( -C P_T^{j2} \right) \right]
\end{equation}

\nn with the constant $C$ properly chosen to reproduce the cross section at
$\cal{O}(\alpha_s^3)$. The obtained result is shown in Fig.~8 at the LHC for
two c.m. energies: $\sqrt{s}=10$ and 14 TeV. The HMRS(B) structure functions
\cite{hmrs} are used and the scale $Q^2$ in the evaluation of the structure
functions and $\alpha_S$ is chosen to be equal to $Q^2=E_T^2=m_Q^2+ P_T^2(Q)$.

The cross section is very large, and one sees that at $\sqrt{s}=14$ TeV,
even for $m_Q \sim 1$
TeV, it is at the  level of 0.1 pb which leads to 1000 events even at a
moderate luminosity of 10 fb$^{-1}$.
The best signature makes use of the ``gold--platted" decay mode where the
heavy quark decays into its light partner and a $Z$ boson, with the latter
subsequently decaying into two charged leptons $l=e,\mu$. This leads to the
spectacular final state
\begin{eqnarray}
pp \ra Q\bar{Q} \ra (qZ) (\bar{q}Z) \ra (jl^+l^-) (jl^+l^-)
\end{eqnarray}
which has a somewhat small branching ratio: $\sim 5. 10^{-4}$ for large quark
masses. Allowing one of the $Z$ bosons to decay into jets or neutrinos, or one
of the $Q$'s to decay into a quark and a W boson (which then decays
into jets or leptons) will drastically enhance the cross section times
branching ratio rate. The obtained signals involve at least two leptons
and are still very interesting. It is therefore very likely that heavy quarks,
with masses up to $\sim 1$ TeV can be found at LHC.

Heavy leptons can also be produced in $pp$ collisions. The processes are: the
Drell-Yan mechanism $q\bar{q} \ra L\bar{L}$ with $\gamma /Z$ exchange for $E$
and only $Z$ exchange for $N$, the gluon-fusion mechanism $gg \ra L\bar{L}$
which proceeds through quark loops and $Z+$ Higgs boson exchange, and for the
charged lepton the $\gamma \gamma$ fusion process $\gamma \gamma \ra L^+L^-X$.
In addition, one has associated production of $N$ and $E$ in the Drell--Yan
process $qq \ra W^* \ra NE $. For singlet neutrinos, since they have
no weak charges, none of these processes is at work; one has therefore
to produce them through mixing and since the angles are small, it is rather
difficult to find these states at $pp$ colliders.
The total cross sections for the charged leptons \cite{S20}
are shown in Fig.~9 at
LHC with $\sqrt{s}=14$ TeV as a function of $m_L$; the ones for neutral
leptons (excluding the $\gamma \gamma$ process) which fully couple to the $Z$
should be of the same order of magnitude. They have been obtained using the
HMRS(B) structure functions at the scale $\mu^2=\hat{s}/4$; $\alpha_S$
is evaluated at the two loop level in the $\overline{MS}$ scheme with
$\Lambda_{\overline{MS}}^4$=0.19 GeV. For the $gg$ fusion, we assumed only
three generations of quarks, and set $m_t=175$ and $M_H=300$ GeV. Fig.~9
shows that the Drell--Yan process has the largest cross section for small
$m_L$, $\sim 1$ pb for $m_L=100$ GeV, but it falls to $\sim 0.1 $ pb for
$m_L=700$ GeV. The $gg$ fusion cross section falls less rapidly and dominates
for $m_L >500$ GeV. Requiring 100 events for $\int {\cal L}=10(100)$ fb$^{-1}$,
one can reach lepton masses of the order of 400 (700) GeV. The lower curves
represent the inelastic (solid), elastic (dashed) and semi--elastic
(dot--dashed) $\gamma \gamma$ cross sections; they are three orders of
magnitude smaller. However, these processes might prove useful in confirming
the signal.

\nn
\begin{figure}[htbp]
\centerline{\psfig{figure=fig7.ps,height=9cm,width=15.5cm,angle=90}}
\vspace*{-5mm}
\caption{\small Total production cross sections (in nb) for heavy quarks at
LHC for two c.m. energy values: $\protect \sqrt{s}=$ 10 and 14 TeV.}
\end{figure}
\vspace*{-5mm}
\nn
\begin{figure}[htbp]
\centerline{\psfig{figure=fig8.ps,height=9cm,width=15.5cm,angle=90}}
\vspace*{-5mm}
\caption{\small Total production cross sections for heavy charged
leptons at LHC with $\protect \sqrt{s}=$ 14 TeV, in the Drell--Yan, the
gluon--fusion and the $\gamma \gamma$ fusion mechanisms.}
\end{figure}

\section{Excited Fermions}

\subsection{Introduction, Lagrangians and Decay Modes}

In this study, we will assume that the excited fermions \cite{R1,R2} have
spin and isospin 1/2 to limit the number of parameters (higher spin and isospin
have been  discussed in \cite{R3}). Furthermore, to accommodate the fact that
the excited  states are much heavier than  the ordinary fermions we will assume
that they  get their masses prior to SU(2)$_L \times $U(1)$_Y$ breaking and
hence, their  couplings to the gauge fields are vector-like.  Therefore,
denoting the excited  fermion doublet by $F^{\star}=F_L^{\star}+ F_R^{\star}$,
the $F^*F^*$--gauge boson interaction  Lagrangian is
\beq
{{\cal L}}_{ f^{\star} f^{\star} } =  \overline{F^{\star}} \; \gamma^{\mu}
\left[ g (\vec{\tau}/2) \vec{W_\mu} \; + \; g' (Y/2) B_{\mu} \;+\;
g_S (\vec{\lambda}/2) \vec{G_\mu} \right] F^{\star}
\eeq
where $\vec{\tau}$ are the Pauli matrices, $Y$ the weak hypercharge ($-1$
for leptons and	$1/3$ for quarks) and	$g,g'$ the usual weak couplings
constants $g=e/\sin \theta_W$ and $g'=e/\cos \theta_W$; $\vec{\lambda}$  are
the Gell-Mann matrices and $g_S$ the strong coupling constant. Note that form
factors and contact interactions may be present, they will be discussed in the
context of hadron colliders where they play an important role.

The Lagrangian describing the transition between excited fermions  and ordinary
fermions should respect a chiral symmetry in order to protect  the light
leptons
from radiatively acquiring a large anomalous magnetic moment \cite{X2}. This
means that only the right-handed part of the excited fermions takes part in the
generalized magnetic de-excitation and  we have \cite{R1,R2}:
\beq
{{\cal L}}_{ f f^{\star} } =  (1/2\Lambda) \overline{F^{\star}} \sigma^{\mu
\nu} \left[ gf (\vec{\tau}/2)\vec{W_{\mu \nu}} +  g'f'(Y/2)B_{\mu \nu} +
g_S f_S (\vec{\lambda}/2) \vec{G_{\mu \nu}} \right] f_{L} + h.c
\eeq
$\Lambda$ is the scale of substructure which we will take of the order of 1
TeV,
while the  $f'$s are weight factors associated with the three gauge groups; the
tensors $V_{\mu \nu}$ are the fully  gauge-invariant  field tensors.  We will
set $f=f'=f_S$: this not only reduces the number of  parameters so that a more
predictive  analysis can be conducted, but also is more natural since for
$f=f'$
the  excited neutrino has no tree-level  electromagnetic couplings
\cite{R1,R2}. Therefore,  apart from the masses of the excited fermions, the
only other parameter is the  strength of the de-excitation $f/\Lambda$.

We will only consider masses for the excited fermions above $M_Z$ since smaller
masses will be probed at LEP2. In this case the two body decays into $W/Z$
bosons and light fermions are kinematically allowed. The decay widths for $f^*
\ra V f$ where $V =\gamma, Z,W$ are given by
\begin{eqnarray}
\Gamma(f^\star \ra f^{(')} V) = \frac{\alpha}{4} \frac{m_{f^*}^3}
{\Lambda^2} f_V^2 \left( 1-\frac{M_V^2}{m_{f^*}^2} \right)^2
\left( 1+\frac{M_V^2}{2m_{f^*}^2} \right)
\end{eqnarray}
with $f_\gamma =e_f f$ , $f_W= f/(\sqrt{2}s_W)$ and $f_Z= (4I^{3}_f- 4
e_fs_W^2)
f/(4s_Wc_W)$.  For excited quarks, there is also the very important decay
$q^\star \ra qg$;  the width is given by the previous equation with $\alpha \ra
4 \alpha_S/3$ and $f_V=f$. The  $f^\star$ have very narrow widths:  for
$m_{f^\star}=500$ GeV and $\Lambda/f$ =1 TeV the total width of the $e^*$ is
less than 1 GeV. For masses much larger than $M_{Z}$, the branching ratios  are
unambiguously predicted since they do not depend on $m_{f^\star}$ and
$\Lambda$; they are given in Table~3.\s

\nn {\small Table 3: Branching ratios for the decays of excited fermions (in
$\%$) for large masses.}

\vspace*{0.3cm}

\centerline{
\begin{tabular}{||c||c|c|c|c||} \hline
$\hspace{0.5cm} f^* \hspace{0.5cm} $  &  $f^* \ra f \gamma $ & $  f^*
\ra f Z  $ &   $  f^* \ra f W $ & $f^* \ra f g$\\ \hline \hline
$\nu^\star$ &0 &  39. & 61. &0. \\ \hline
$e^{\star} $& 29. &  11. & 60. &0. \\ \hline
$u^{\star}$ &2. & 3. & 10. & 85.\\ \hline
$d^{\star}$ &0.5 & 5. & 10. & 84.5 \\ \hline
\end{tabular}}
\vspace*{0.3cm}

As one can see, the electromagnetic decay of charged excited fermions is not
the
dominant one. For $e^*$, it is just about $30\%$ compared to almost $100\%$
for
masses smaller than $M_W$.  For quarks, this electromagnetic decay, which
would
constitute the cleanest way for ``tracking" these particles, is a very  small
fraction of all decays; therefore, relying on this mode leads to a
considerable
loss of events. Nevertheless, it constitutes a very characteristic signature of
excited fermions and could help to disentangle them from the exotic fermions
discussed previously.

\subsection{Production at $\ee$ Colliders}

\underline{Pair Production.} If kinematically allowed, excited fermions can be
pair-produced without any  suppression due to the factor $f/\Lambda$ (form
factors might be present, though). In $e^{+}e^{-}$  collisions the reaction
proceeds through $\gamma$ and $Z$ $s$--channel exchange  for charged fermions,
whereas for excited neutrinos there is only a $Z$  exchange for $f=f'$; the
charged excited fermions  can also be pair produced in the  $\gamma \gamma$
mode
of the collider. The differential and total cross sections are the same as for
the $vector$-$like$ exotic fermions discussed previously. The only difference
will be
in the decay modes: while exotic fermions will decay only to $W/Z$ and light
fermions, excited charged fermions have the  electromagnetic decay and excited
quarks will dominantly decay into quarks plus gluons; see Tab.~3. Since the
production rates are rather large (see the previous discussion on exotic
fermions, in particular Figs.~1), all these final states can be easily
searched
for in the  clean environment of $\ee$ colliders, and the discovery limits that
can be reached will be nearly
the kinematical limit of $m_{f^*} \sim \sqrt{s}/2$ \cite{R2}. \s

\nn \underline{Single Production.} Owing to the special coupling of the excited
fermions to their light partners, one can also have single production of the
excited particles. Hence, in  principle, $f^*$ masses up to the total c.m.
energy of the collider can be  probed. However, the rates depend on the
parameter $f/\Lambda$ which measures  the strength of the transition. We will
consider single production at  $\ee$ colliders in the three modes: $\ee,
e^-\gamma, \gamma \gamma$. We will take $\Lambda/f=$~1 TeV in  our numerical
analysis.

In \underline{$\ee$ annihilation}, the single production proceeds through
$s$--channel $\gamma$ and $Z$ exchange for all excited fermions. For the first
generation of excited leptons, one has substantial contributions due to
additional $t$--channel diagrams: $W$ exchange in the case of the $\nu^*$; $Z$
and the important $\gamma$ exchange in the case of the $e^*$. These processes
should be compared to the single production of exotic heavy fermions; however,
in the latter case, there is no photon exchange and the couplings are not of
the
magnetic type. The total section for the single  production of $\nu^*_e$, which
is the same for $ \overline{\nu_{e}} \nu^\star$ is \cite{R2}
\begin{eqnarray}
\sigma & = &\sigma_0 \frac{3s}{4 \Lambda^2} \beta  \left\{(1- \frac{2}{3}
\beta)
\beta A_\nu + \frac{f_W^2}{2s_W^2} \left[ (2w+ \beta ) \log(1+\frac{ \beta }
{w}) -2 \beta \right] \right. \nonumber \\
& + & \left. \frac{2f_W}{\sqrt{2}s_W} \left ( e_e f_\gamma +\frac{a_e+v_e}{1-z}
f_Z \right) \left[w(1+\frac{w}{ \beta }) \log(1+\frac{ \beta }{w}) -\frac{1}{2}
 ( \beta +2w) \right] \right\}
\end{eqnarray}
where $\beta=(1-m_{\nu^*}^2/s)$, $w=M_W^2/s$, $z=M_Z^2/s$ and $A_f$ (with
$f=\nu$) is given by
\beq
A_f = e_{e}^2 e_f^2 + \frac{2e_{e} v_e f_\gamma f_Z }{1-z}
+\frac{(a_e^2+v_e^2)f_Z^2} {(1-z)^2}
\eeq
These formulae may be used for all other flavours (except for the  $e^{\star}$)
by setting $f_W=0$ and by including the colour factor for quarks.  For the
$e^\star$, the expression of the total cross section is quite  involved
and can be found in \cite{R2}; a  very good approximation is to consider  only
the $s$ and $t$ channel photon exchange, where the much simpler  expression is
given by \cite{R2}
\begin{eqnarray}
\sigma = \sigma_0 \frac{3s}{4 \Lambda^2} f_\gamma^2 \ \beta \left[ (1- \frac{2}
{3}\beta) \beta -\beta -3  +\frac{1+\beta^2}{\beta} \log \left(\frac{s}{m_e^2}
\frac{\beta^2}{(1-\beta)^2} \right) \right]
\end{eqnarray}
In Figs.~10a/b we show the total cross sections at a c.m. energy of 1 TeV.
The largest production rate  occurs for the $e^\star$ due to the
$t$-channel photon exchange: compared to the other charged leptons this has a
two-order of magnitude enhancement. The same is true for the $\nu_e^\star$
as compared to the other excited neutrinos as a result of the $W$ exchange.
Charged excited fermions should be looked for by exploiting their
electromagnetic decays; requiring a cut on the transverse momentum of the
photon to be larger than $\sim$ 20 GeV together with a rapidity cut of $\sim$
2 should be sufficient to suppress potential backgrounds
from radiative QED processes. For the $e^*$, slightly more severe cuts should
be applied to further reduce the Bhabha background; one can also use the
$e^+ e^- jj$ final states, similarly to what has been discussed in the
case of heavy charged leptons. For excited neutrinos, the situation is also
similar to that of exotic neutrinos (although the distributions are different)
and one has to look for $e \nu jj$ events. A detailed analysis of the
background has not been performed here; requiring 20 events to establish a
signal for $\int {\cal L}=$ 100 fb$^{-1}$, one would reach masses close to the
c.m. energy for first generation leptons and slightly smaller masses for the
other fermions.

\nn
\vspace*{-5mm}
\begin{figure}[htbp]
\hspace*{-1cm}
\centerline{
\psfig{figure=fig9.ps,height=21cm,width=15cm}}
\vspace*{-1.7cm}
\caption{\small Cross sections for the single production of excited fermions
in $\ee$ collisions with $\protect \sqrt{s}=1$ TeV (a,b), of excited neutrinos
in $e\gamma$ collisions (c) and of excited charged fermions in $\gamma \gamma$
collisions (d).}
\end{figure}

\newpage

In \underline{$e \gamma$ collisions}, excited leptons of the first generation
can be also singly produced. In fact, a motivation for running in this mode
would be the production of $e^{\star}$ as a resonance, thus (if this particle
exists) turning the machine into an $e^{\star}$ factory. The cross-section
after integrating over the Breit-Wigner resonance is\cite{R4}
\begin{eqnarray}
\sigma( \gamma e \ra e^{\star} )\;=\; \frac{8 \pi^2}{m_{e^\star}^{2}}
\frac{\Gamma(e^{\star} \ra e \gamma )}{m_{e^\star}}
\end{eqnarray}
The best channel to detect this particle is the electromagnetic decay, although
there is a potential large background from Compton scattering. However, the
bulk of these background events is along the beam direction. Moreover, the
produced electron from Compton scattering flies opposite to the initial
electron from the beam whereas the electron from the signal has a spherical
distribution. Requiring an observation of $20$ events over the background $e
\gamma \ra e \gamma$, one can reach a limit on the scale $\Lambda$ of about
$200$ TeV, provided that the mass is below the kinematical limit which is
approximately $900$ GeV for a $1$ TeV $e^{+} e^{-}$ collider.

In the $e\gamma$ mode, one can also search for the $\nu_e^\star$ which can be
produced in association with a $W$, $e^- \gamma \ra \nu_e W^-$. Therefore one
can in principle reach masses of the order of 800 GeV. Fig.~10c shows the total
cross section for various values of $m_{e^*}$ (which can be exchanged in the
$s$--channel) and even for infinite $m_{e^*}$, where it is the smallest, the
cross section is larger than in $\ee$ collisions. For the signature, one can
use the decay into $e^{-} W^{+}$ and by letting both the primary $W^-$ and the
decay $W^{+}$ go into jets which seems to be background-free. The bounds on
$\Lambda$ which can be reached when requiring $20$ events, are ${\cal O}(100$
TeV) for $m_{\nu_e^*}$ smaller than $\sim 800$ GeV, assuming the same
luminosity as in the $\ee$ mode. \s

\underline{In $\gamma \gamma$ collisions}, all charged excited fermions can be
singly produced through two $t$--channel exchanges: one involving $f^*$ and the
other $f$. The latter gives a very large contribution, similarly to the
$t$--channel enhancement  with $e^{\star}$ production in $\ee$ collisions. The
differential cross-section is forward/backward peaked, an
effect which is more pronounced for the lightest ordinary fermions; this mass
effect however disappears when we keep to scattering angles $|\cos \theta|\leq
0.8$. In terms of $a=m_{f^*}^2/s$, the total cross section for $f^*$ production
is \cite{R2}
\begin{small}
\begin{equation}
\sigma=\sigma_0 N_c e_f^4 f^2 \frac{6s}{\Lambda^2} \left[ a(2a^2-2a+1)\log
\frac{(1-a)^2}{m_f^2/s}+ \frac{a(1-a-2a^3)}{1+a} \log a +(1-a)
(3+4a^2) \right]
\end{equation}
\end{small}
Even in the case where the particles are produced at small angles, which
accounts for a large part of the cross section, the events are not lost since
the decay products of the $f^\star$ are at large angles. As Fig.~10d shows,
$\sigma$ increases with $m_{f^\star}$ up to nearly the kinematical limit where
it starts bending over. The importance of the $\gamma \gamma$ mode compared to
$\ee$  is that single production (even for $d$--type quarks where the charge
is penalizing, and for a cut $|\cos\theta|\leq 0.8$) is larger for all flavours
except for the $e^*$ with $m_{e^\star}<$ 700 GeV. The signals are quite clean:
for $q^\star$, one can rely on the dominant decay $q^\star \ra qg$, where the
very energetic $q$ and $g$ jets are emitted at large angles (two hard QCD
jet-events can be eliminated by imposing the cut $|\cos \theta| <0.8$)
while for $e^*$ one can use the $e^* \ra e \gamma$ mode. Therefore, one can
probe excited fermion masses of the order of 800 GeV, for reasonable
$\Lambda$ values.

\subsection{Production in $eP$ Collisions}

Due to the special couplings of the electron to the excited leptons of the
first generation, one can have single production of $e^*$ through $t$-channel
$\gamma$ and $Z$ exchange, and $\nu^*$ through $t$-channel $W$ exchange in
$eP$ collisions. Excited quarks of the first generation can also be produced
in the same way, however background problems make this possibility less
interesting than the production of excited leptons to which we stick here
\cite{R5}.
Using the scaled mass $a =m_{f^\star}^2/\hat{s}$, the deep inelastic
differential
cross section for the process $eP \ra l^* X$ reads \cite{R2}
\begin{eqnarray}
\frac{\sigma}{dx dy}=2\pi \alpha^2 \frac{\hat{s}^2}{\Lambda^2}
 y \ \sum_{q,\bar{q}}\left[ A_l(Q^2)R(x,y)q(x,Q^2)+\bar{A}_l(Q^2)\bar{R}(x,y)
\bar{q}(x,Q^2)	\right]
\end{eqnarray}
\vspace*{-3mm}
\begin{eqnarray}
with \ \ \ R(x,y)  = 2-(2-a)(y+a) \ , \ \ \ \
\bar{R}(x,y) =	a( 2-y-a)
\end{eqnarray}
and in terms of	the quark couplings to the gauge bosons, $A_l$ and
$\bar{A_l}$ are	defined	by
\begin{small}
\begin{eqnarray}
A_\nu =	\bar{A}_\nu = \frac{f_W^2}{4s_W^2} \frac{1}{(Q^2+M_W^2)^2}
\nonumber
\end{eqnarray}
\vspace*{-0.6cm}
\begin{eqnarray}
A_e = \frac{e_q^2 f_\gamma^2}{(Q^2)^2}+	\frac{2e_q v_q f_\gamma	f_Z}
{Q^2(Q^2+
M_Z^2)}+\frac{(v_q^2+a_q^2)f_Z^2}{(Q^2+M_Z^2)^2} \   \hspace{0.5cm}
\bar{A}_e = \frac{2e_q a_q f_\gamma f_Z}{Q^2(Q^2+ M_Z^2)}+
\frac{2	v_q a_q	f_Z^2}{(Q^2+M_Z^2)^2}
\end{eqnarray}
\end{small}
In addition to the previous contribution (with a $Q^2$ cut of 5
GeV$^2$), one has two other contributions for the $e^*$: one due to low
$Q^2$ deep inelastic scattering	and another due	to the elastic process
$eP \ra	e^*P$. The integrated total	cross sections
for $e^*$ and $\nu^*$ production are shown in Fig.~11a for $\Lambda/f=1$
TeV. For $e^*$ the three different contributions discussed above are
shown separately. Due to the low $Q^2$ $t$-channel photon exchange, the $e^*$
total cross section is about an	order of magnitude larger than for
$\nu^*$. A clean  detection channel  will be provided by the wide angle
electron-photon	pair final state in the	case of	the $e^*$, and the
electron--$W$ final state in the case of the $\nu^*$. Requiring	20
such events and	assuming an integrated luminosity of 1 fb$^{-1}$,
masses up to 800 GeV for $\nu^*$ and $e^*$ can be probed.

The excited leptons (for $e^*$ those produced in the deep inelastic process)
have larger $P_T$ than exotic leptons; Fig.~11b. This feature, in addition
to the different $y$ distributions, can help to disentangle between the two
sorts of new leptons which have the same decay modes and branching ratios
(for the new electrons the distinction can easily be made because the $e^\star$
can decay into a photon). An additional way to disentangle between the two
different
sorts of neutral heavy leptons, is the completely different final polarization
as shown in Fig.~11c for the $l^*$ (it is almost the same for $e^*$ and
$\nu^*_e$).
\nn
\begin{figure}[htbp]
\hspace*{-1cm}
\centerline{
\psfig{figure=fig10.ps,height=20.5cm,width=15cm}}
\vspace*{-1.7cm}
\caption{\small Total cross sections for the single production of first
generation excited leptons in $eP$ collisions with $\protect \sqrt{s}=1.2$ TeV
(a); the transverse momentum distribution (b) and the longitudinal and
transverse polarizations (c).}
\end{figure}

\subsection{Production in $pp$ Collisions}

Excited quarks can be produced in $pp$ collisions through a variety of
mechanisms \cite{R6,S18}. The dominant production channels are the gluonic
excitation of  quarks $g + q \rightarrow q^{*}$ which occurs through the
$q^*qg$
``gauge" interaction described previously, and the excitation through preon
interactions $qq  \rightarrow qq^{*}$ and $q^{*}q^{*}$; through contact
interactions excited  leptons, too, could eventually be produced at observable
rates, $q\bar{q}  \rightarrow ee^{*}$ and $e^{*}e^{*}$. The signatures of
excited quarks are  bumps in the invariant energies of jets, jet + gauge boson
and jet + lepton  pair combinations. Excited leptons would reveal themselves in
bumps of leptons, leptons + gauge particles or leptons + quark jets. The first
indication for  the production of novel excited fermions could be the copious
production of  leptons, at large rates unexpected in the framework of the
Standard Model.

The cross section for the gluonic excitation of quarks $gq\to q^*$
in $pp$ colliders is given by (we have taken $\Lambda=m_*$ in the
numerical analysis) \cite{R6}
\begin{equation}
\sigma = \frac{\alpha_{s} \pi^{2}}{3\Lambda^{2}} \tau \frac{d{\cal L}^{gq}}{d
\tau} \ \ \ \ ,  \ \ \ \ \ \tau=\frac{m_{*}^{2}}{s}
\end{equation}
where $d{\cal L}^{gq}/d\tau$ is the quark--gluon luminosity for the $pp$ beams.
The production cross section is shown by the full line in Fig.~12a, for the LHC
at $\sqrt{s}=14$ TeV. Given an integrated luminosity of 10 fb$^{-1}$, a mass
range of 5--6 TeV can be reached in this channel, based on 100 to 1000 events.
The signals for singly produced excited quarks are large transverse momentum
$jj$, $j\gamma$, $jZ$ or $jW$ pairs peaking at the mass of the resonance.
Because the final states of the signal consist of large $P_T$ jets with
large angles $\theta_{jj}$ between the jets of each pair, we introduced
the following cuts to reduce the background:
$$\theta_{jj} > 30^{\circ} \ \ , \ \ P_T > 100~GeV \ \ ,  \ \
\eta < 2.5  \ \  , \ \ E^{tr}   > m_*/2 $$
The mass resolution is determined by the decay width of the resonance
and the experimental jet resolution, which is taken to be $\Delta E/E = 0.35/
\sqrt{E} + 0.02$.

Excited quarks of the first generation can also be produced via contact
interactions (which for large masses can overwhelm the gauge interactions) in
the processes $qq \rightarrow  qq^{*}$ and $qq \ra q^* q^*$; for the value
$\Lambda=m_*$ one can reach  $q^\star$ masses of the order of 6 TeV and 4 TeV
respectively; Fig.~12b. The backgrounds, which together with the cross
sections
have been calculated in \cite{R6},  are well under control as shown in the
figure.   Through contact interactions, excited leptons could also be produced
copiously   in the processes $q\bar{q}\rightarrow ee^{*}$ and $q\bar{q}
\rightarrow e^*e^*$. The cross sections, which are shown in Fig.~12c,  are
large
and the signals, consisting of pure leptonic channels, would provide  very
clear
signatures for the experimental identification of these novel states [the
backgrounds are very rare in the SM]. For a luminosity of 10
fb$^{-1}$ $e^*$ masses up to $\sim 4$ TeV could be accessible
for $\Lambda=m_*$ \cite{S18}.
\begin{figure}[htbp]
\centerline{\psfig{figure=fig11a.ps,height=9.5cm,width=15.5cm,angle=-90}}
\centerline{\psfig{figure=fig11b.ps,height=9.5cm,width=7.5cm,angle=-90}
            \psfig{figure=fig11c.ps,height=9.5cm,width=7.5cm,angle=-90}}
\vspace*{-0.6cm}
\caption{\small Total cross sections for the production of (first
generation) excited quarks through gauge interactions (a) or contact
interactions (b) and of excited leptons through contact interactions in $pp$
collisions with $\protect \sqrt{s}=14$ TeV.}
\end{figure}

\section{Difermions}

\subsection{Leptoquark Production at $e^+e^-$ Colliders}

There is much interest in the study of leptoquarks (LQ), colour (anti-)triplet,
spin--0 or spin--1 particles, which carry both baryon and lepton quantum
numbers.  As discussed in the introduction, such objects appear in a large
number of extensions of the SM such as GUT's,  Technicolour, and composite
models.  Quite generally, the signature for  leptoquarks is striking: a high
$p_t$ lepton balanced by a jet, or missing  $p_t$ balanced by a jet, for the
$\nu q$ decay mode, if applicable.

Single and pair production of scalar LQ's at a linear $e^+e^-$ collider of
$\sqrt{s}=1$ TeV was summarized by {\cite {BLN}}. All three modes of the
collider -- $e^+e^-$, $e\gamma$ and $\gamma\gamma$ -- were analyzed.  We
consider first the quark-level contribution to the process $e^-\gamma\to qS$,
where $S$ is the LQ \cite{BLN}. (This process was first considered by  {\cite
{orig2}}.) We parameterize the strength of the LQ coupling by
comparing it to the electromagnetic interaction, i.e., $g_{\sss LQ}^2=4\pi
k\alpha_{em}$, and allow $k$ to vary. The cross section is
\beq
\label{LQsigma}
\sigma(s) = {\pi k \alpha_{em}^2 \over 2 s}
(\qs+1)^2 (1-2\alpha+2\alpha^2) \ln\left[{s\over 4 m_q^2}
\left(\alpha+\beta\right)^2 \right] + ...
\eeq
where `...' indicates additional (subdominant) terms, $\qs$ is the LQ charge
and
\beq
\alpha\equiv 1 - (\ms^2 - m_q^2)/s~,~~
\beta^2 \equiv  1 - 2(\ms^2+m_q^2)/ s + (\ms^2-m_q^2)^2/ s^2
\eeq
Note that, due to the factor $(\qs+1)^2$ in eq.~\ref{LQsigma}, the
production cross sections for LQ's of charge $-5/3$ and $-1/3$ are equal,
as are those for $Q_S=-4/3$ and $-2/3$ (up to subdominant terms). The
apparent divergence in the case of massless quarks is removed when detector
cuts are imposed.

There are additional
contributions to LQ production due to the hadronic content of the photon
\cite{DG}. These can be taken into account by using a resolved photon,
i.e.\ a photon distribution function. In Fig.~13 we compare the two
contributions for $\sqrt{s}_{ee}=1$ TeV, for $\qs=-5/3$ and $k=1$, using
the GRV distribution functions \cite{GRV} with $Q^2=\ms^2$. In this figure
we have folded in the photon energy spectrum due to the backscattered laser
light. One sees that the resolved photon contribution is larger than the
direct contribution for all LQ masses. This is easily understood
physically: when one uses a resolved photon, one is actually considering
the process $e^-\gamma\to S X$, where $X$ is not simply $q$ (as was the
case above), but rather includes all sorts of soft hadronic products. Since
more final states are included, relative to the direct process, it is only
natural that the cross section should be larger. It should be pointed out,
however, that, except for very light LQ's, the enhancement is only a factor
of 2-3. In what follows we will consider {\it only} the resolved photon
contributions to the different processes.

The dominant contribution to single LQ production in $e^+e^-$ collisions
comes from the sub-process $e^-\gamma\to Sq$, in which the photon is
radiated from the $e^+$. The cross section for $e^+ e^- \to e^+ S q$
is then calculated using the effective photon approximation. In addition,
an $e^+e^-$ collider can be turned into an $e\gamma$ or $\gamma\gamma$
collider by backscattering laser light from one or both of the beams. In
calculating cross sections for processes at such colliders, we take into
account the energy spectrum of the backscattered photons.

\vspace*{-1cm}

\begin{figure}[htbp]
\centerline{\psfig{figure=eglog.ps,height=7.5cm}}
\vskip-1.1truecm
\caption{Single LQ production cross sections for direct (`large log') and
resolved photon contributions.}
\end{figure}

In Fig.~14 we compare the single-LQ production cross sections for all LQ
charges, for $k=1$, at $e^+e^-$, $e\gamma$ and $\gamma\gamma$ colliders.
There are several features to these figures. First, just as was the case
for the direct contributions, the cross sections for LQ's of charge $-5/3$
and $-1/3$ are equal, and similarly for $Q_S=-4/3$ and $-2/3$.
Second, at $e^+e^-$ and $e\gamma$ colliders, only
those LQ's which couple to the first generation ($eu$ or $ed$ LQ's) can be
produced, but LQ's of all three generations can be produced at
$\gamma\gamma$ colliders. Note that, since the $t$-quark distribution
function is unknown, for the third-generation LQ with $Q_S=-5/3$ we
calculated the direct contribution only.

As a figure of merit, we assume a luminosity of 60
fb$^{-1}$, and require 25 events for discovery. This implies that a LQ is
observable if its production cross section is larger than 0.4 fb. It is
clear from Fig.~15 that the $e\gamma$ mode is the best way to look for LQ's.
For all charges, LQ's of mass up to about 900 GeV are observable. For
$Q_S=-5/3$, one can go slightly beyond this limit in the $e^+e^-$ mode.
$\gamma\gamma$ colliders are clearly not competitive for first-generation
LQ's. However, second- and third-generation LQ's are visible for certain
ranges of masses.  For other coupling, one simply scales the curves linearly
in $k$. Thus, at $e\gamma$ colliders, first-generation LQ's will be observable
even for couplings as weak as $k\lsim 10^{-2}$-$10^{-3}$.

For all three colliders, the bulk of the cross section comes from those
processes in which all particles go directly down the beam pipe. However,
the LQ will subsequently decay, and its decay products will be seen in the
detector. The signal will therefore simply consist of a lepton and a jet with
a negligible SM background.

\newpage

\vskip-0.6truecm

\begin{figure}[htbp]
\centerline{\psfig{figure=eeeggg.ps,height=7cm}}
\vskip-0.6truecm
\caption{Single LQ production cross sections at $e^+e^-$, $e\gamma$ and
$\gamma\gamma$ colliders for a) $\qs=-5/3$ or $-1/3$, b) $\qs=-4/3$ or
$-2/3$.}
\end{figure}
\vskip-1.5truecm
\begin{figure}[htbp]
\centerline{\psfig{figure=pair.ps,height=7cm}}
\vskip-0.6truecm
\caption{LQ pair-production cross sections at $e^+e^-$ and $\gamma\gamma$
colliders.}
\end{figure}
\vskip-.1truecm

For very weak LQ couplings, a better signal rate may be obtained from
pair production of LQ's at $e^+e^-$ or $\gamma\gamma$ colliders. Of course,
the search is limited to LQ's of mass less than half of the c.m. energy of
the collision. The cross sections for such a process are
presented in Fig.~15. At $e^+e^-$ colliders, there are two contributing
diagrams, one of which depends on the LQ coupling $k$. The other diagram
depends on the LQ's charge and weak isospin (in what follows, we assume
$I_3=-1/2$ for LQ's with $Q_S=-5/3$ and $I_3=0$ for $\qs=-1/3$).

At $\gamma\gamma$ colliders, on the other hand, the cross section is
$k$-independent, depending solely on $Q_S$. In Fig.~15a,
we show the pair-production cross sections for a LQ of charge $-5/3$. At
$e^+e^-$ colliders, we see that one loses about an order of magnitude in
cross section as one passes from $k=1$ to $k=0$. However, regardless of the
coupling, pair production of LQ's is observable for LQ masses essentially
up to $\sqrt{s}/2$. For comparison we also show the single-LQ production
cross section. Both single- and pair-production cross sections are roughly
of the same size for $k=1$, within the given LQ mass range, but the
pair-production cross section wins out for smaller values of $k$.
(Note that, in pair production, there may be more SM background). We
also show the cross section at $\gamma\gamma$ colliders, including both the
$\gamma\gamma$- and $\gamma g$-initiated pair production. Unless $k$ is
significantly bigger than 1, the LQ pair-production rate is greater at
$\gamma\gamma$ colliders than at $e^+e^-$ colliders for the smaller values
of the LQ mass. In Fig.~15b we present the cross sections for the smallest
value of the LQ charge, $Q_S=-1/3$. Here we see that LQ pair production
proceeds at a much greater rate at $e^+e^-$ colliders, for $k=1$. For
$k=0$, on the other hand, the $\gamma\gamma$ collider is better for smaller
LQ masses. Again, LQ's of any generation are observable in
this process for masses up to a bit less than $\sqrt{s}/2$.

We conclude that it will be necessary to use all three modes of the linear
$e^+e^-$ collider, and to consider both single and pair production, in order to
perform a complete search for LQ's.

Although the discovery of a
leptoquark would be dramatic evidence for physics beyond the SM, it
would lead to the question of which model the leptoquark originated from.
Given the large number of leptoquark types it would be imperative to measure
its properties to answer this question.
Following the notation of Ref.~\cite{buch},
the complete set of possible LQ are: $S_1$, $\tilde{S}_1$ (scalar,
iso-singlet); $R_2$, $\tilde{R}_2$ (scalar, iso-doublet); $S_3$ (scalar,
iso-triplet); $U_1$, $\tilde{U}_1$ (vector, iso-singlet); $V_2$, $\tilde{V}_2$
(vector, iso-doublet); $U_3$ (vector, iso-triplet).  The production and
corresponding decay signatures are quite similar, though not identical, and
have been extensively studied. Even focussing only on the NLC
($e^+ e^-$, $e \gamma$ and $\gamma \gamma$ modes), there is a
already a considerable amount of work in the literature {\cite{leptoee}}.
The possibility of using a polarized
$e\gamma$ collider to differentiate the LQ type (\ie, a polarized $e$ beam
in conjunction with a polarized-laser backscattered photon beam) was analyzed
here {\cite {donc}} including the contributions due to the
hadronic content of the photon. An integrated luminosity of 50 fb$^{-1}$ was
assumed.

\vspace*{-0.5cm}
\nn
\begin{figure}[htbp]
\centerline{
\psfig{figure=mikea.ps,height=6cm,width=8cm,angle=90}
\hspace*{-5mm}
\psfig{figure=mikeb.ps,height=6cm,width=8cm,angle=90}}
\vspace*{0.01cm}
\caption{\small $A_{LL}$ {\it vs.} $M$ for (a) LQs which couple only to LH
electrons for a 1~TeV collider.  The (solid, dashed, dotted, dot-dashed) curves
are for ($S_3$, $\tilde{V}_2$, $U_3$, $\tilde{R}_2$) LQs. (b) Same as (a) but
for LQs which couple only to both LH and RH electrons; here $\kappa_L = 1=
\kappa_R$.  The (solid, dashed, dotted, dot-dashed) curves are for ($S_1$,
$V_2$, $U_1$, $R_2$) LQs.}
\end{figure}

Table~2 of Ref.~\cite{buch} gives information on the couplings to
various quark and lepton combinations; note that both the quark and
lepton have the same helicity (RR or LL) for
scalar LQ production while they have opposite helicity (RL or LR) for vector
LQ production.  Denoting the various helicity cross sections as
$\sigma^{\lambda_e \lambda_q}$, $\lambda_i = +$ for R helicity, $\lambda_i = -$
for L helicity and $\sigma_{TOT} = \sigma^{++} + \sigma^{+-} + \sigma^{-+} +
\sigma^{--}$, it is useful to introduce the double longitudinal spin
asymmetry $A_{LL}$:
\begin{equation}
A_{LL} = \frac{(\sigma^{++} + \sigma^{--}) - (\sigma^{+-} + \sigma^{-+})}
{(\sigma^{++} + \sigma^{--}) + (\sigma^{+-} + \sigma^{-+})}
\end{equation}
which can be calculated for each LQ species. $100\%$ polarization of both
beams is assumed. In the analysis {\cite {donc}}, the {\it asymptotic}
polarized
photon distribution functions {\cite{polpho}} were used, and it is assumed that
$Q^2$ and $x$ are large enough that the Vector Meson Dominance part of the
photon structure is not important, but rather the behavior is dominated by the
point-like $\gamma q \bar{q}$ coupling.

LQ's were identified only in the $ej$ mode. LQ's of spin-0(1) were found
to have positive(negative) values of $A_{LL}$; for LQ's which couple in a
RH manner to $e$'s, this is sufficient to separate all cases. For the case
of a 1 TeV collider, Figs.~16a-b show $A_{LL}$ as a function of the LQ mass
including statistical errors for several LQ species. When LH
couplings are present, then the values of $A_{LL}$ must be carefully
examined. With statistical errors only, an NLC working in the $\gamma e$ mode
could separate all LQ types up to approximately $80\%$ of the center of mass
energy assuming LQ Yukawa couplings of
order electromagnetic strength. The largest uncertainty in the
calculation is the reliability of the asymptotic approximation used to
determine the photon distribution functions and those associated with the
quark content of polarized photons.

The effects of QED and QCD corrections on the production of both vector (V)
and scalar (S) LQ pairs in $e^+e^-$ collisions at the NLC
were considered {\cite {blum}}. These corrections were found to be
critical, if one is to differentiate the different LQ species, since
they significantly modify total cross sections and asymmetries. The vector LQ's
were assumed to have {\it minimal} gauge boson couplings in this analysis so
that the possibility that vector LQ are gauge particles was not covered.
Initial state QED corrections
were performed using the structure function approach including terms up to
order $\alpha^2$ with soft photons exponentiation. Beamstrahlung corrections
were also taken into account.
The full set of QED corrections were found to be only
weakly sensitive to variations in $\sqrt {s}$ in the range 0.5-1 TeV. However,
bremsstrahlung corrections were found to be quite important in the threshold
region, being of order $30-50\%$ depending on the LQ spin.

For the scalar  case, the QCD final state corrections were large and positive
increasing  dramatically near threshold indicating possible bound state
formation. The   QCD corrections were always in excess of $40\%$ and determined
to be more  than $100\%$ for LQ velocities below 0.2 due to the possible
production of  bound states and the well known Coulomb singularity.

\vspace*{-0.5cm}
\nn
\begin{figure}[htbp]
\centerline{
\psfig{figure=joa1.ps,height=8cm,width=8cm,angle=90}
\hspace*{-5mm}
\psfig{figure=joa2.ps,height=8cm,width=8cm,angle=90}}
\vspace*{-1cm}
\caption{\small $95\%$ C.L. discovery region (to the left of the curves) in the
leptoquark coupling--mass plane for (a) LEP II (200 GeV)
and an integrated luminosity of $100, 200, 500$ pb$^{-1}$ corresponding
to the dashed, dash-dotted, and solid curves, respectively.  (b) NLC with a
center of mass energy of 0.5(1.0) TeV and $50 (100)$ pb$^{-1}$, corresponding
to the solid (dashed) curves, respectively.}
\end{figure}

Leptoquarks can participate in the process $e^+e^-\to q\bar q$ via virtual
$u$- or $t$-channel exchanges and can produce deviations from the SM
predictions
for cross sections and asymmetries \cite{oldlq}; thus indirect limits on LQ
properties can be obtained. If one allows for the general form of
leptoquark-fermion interactions of ${\cal L}=\ell (A+B\gamma_5)q\cdot $LQ,
the leptoquark couplings can be parameterized in terms of two constants,
$\kappa\equiv (|A|^2+|B|^2)/e^2$ and $\kappa'=2\Re (A^*B/e^2)$.
Here we examine the case $\kappa'=\kappa$ and limit our discussion to the
exchange of charge $-1/3$ leptoquarks present in $E_6$ theories,
which mediate the reactions $e^+e^-\to u\bar u,c\bar c,t\bar t$.

The results are not found to be qualitatively different for $Q=+2/3$ leptoquark
exchange or for considering the other extreme case of the parameters
$\kappa'=-\kappa$.  The $95\% $ C.L. discovery reach in the leptoquark
coupling--mass plane is presented in Fig. 17 for LEP II with $\sqrt s=200$ GeV,
and the NLC for $\sqrt s=0.5, 1.0$ TeV.  These search regions are obtained via
a
combined analysis of production cross sections, forward-backward asymmetries,
and left-right polarization asymmetries (for the NLC only), assuming a beam
polarization of $90\%$ and a $50\%$ efficiency for the identification of final
state charm particles. Clearly, this process offers a good tool in the
exploration of indirect leptoquark effects.

\subsection{Leptoquark Production at Hadron Colliders}

The calculations for the production of scalar and vector LQ's at hadron
colliders, both singly and in pairs has been updated {\cite {huge}} for this
report. The results in the scalar case were obtained long ago {\cite {handp}},
so we briefly discuss the intricacies of the vector case below.

In order to determine the $q\bar q,~gg \to VV$ cross sections we need to
determine both  the trilinear $gVV$ and quartic $ggVV$ couplings, which may
naively at first  appear to be unknown. However, in any realistic model wherein
vector  leptoquarks appear and are fundamental objects, they will be  the gauge
bosons of an extended gauge group. In this case the $gVV$ and  $ggVV$ couplings
are completely fixed by gauge invariance.  These  particular couplings will
also
insure that the subprocess cross section obeys  tree-level unitarity, as is the
hallmark of all gauge theories.  Of course, it might be that the appearance of
vector  leptoquarks is simply some low energy manifestation of a  more
fundamental theory at a higher scale and that these particles may even  be
composite, in which case so-called `anomalous' couplings in both the $gVV$  and
$ggVV$ vertices can appear. One such possible coupling is an `anomalous
magnetic moment', usually described in the literature by the parameter
$\kappa$
{\cite {hagi}}, which takes the value of unity in the  gauge theory case.

%
\nn
\begin{figure}[htbp]
\centerline{
\psfig{figure=vlq1a.ps,height=9cm,width=8cm,angle=90}
\hspace*{-5mm}
\psfig{figure=vlq1b.ps,height=9cm,width=8cm,angle=90}}
\vspace*{-1cm}
\caption{\small Production cross section for a pair of vector leptoquarks at
the Tevatron: (a) as a function of the LQ mass with $\kappa=1$. The
dotted(dashed, solid)curve corresponds to the $q \bar q$($gg$, total)
contribution. The dash-dotted curve is the total S-LQ result.
(b$ \kappa$ dependence of the $q \bar q$
(dots), $gg$(dashes), and total(solid) $V$ pair production cross sections
at the Tevatron for a vector leptoquark mass of 200 GeV. }
\end{figure}
%
\nn
\begin{figure}[htbp]
\centerline{
\psfig{figure=vlq2a.ps,height=9.5cm,width=8cm,angle=90}
\hspace*{-5mm}
\psfig{figure=vlq2b.ps,height=9.5cm,width=8cm,angle=90}}
\vspace*{-1cm}
\caption{\small Same as the previous figure but for the LHC.
In (b), a vector leptoquark mass of 1 TeV is assumed.}
\end{figure}
\vspace*{1mm}

\noindent Among
these `anomalous couplings', the term which induces $\kappa$ is special in that
it is the only one that conserves $CP$ and is of dimension 4. As  values of
$\kappa$ differing from one have been entertained in the literature,  we will
generally assume $\kappa=1$ or 0, with the latter  value corresponding to
`minimal' coupling, in order to probe the sensitivity  of our results to the
assumed gauge nature of $V$. We will also describe the  results in the more
general case where $\kappa$ is arbitrary.

If LQ's are first observed at hadron colliders it will become necessary to  be
able identify which one {\cite {buch}} has been found. The simplest probe  of
LQ
properties is the cross section itself. The two individual subprocess result in
the total cross sections displayed in Figs. 18a-b and 19a-b at the Tevatron and
LHC and are  compared with the scalar LQ case.
As we see from these figures,
the production rate for spin-1 LQs can be substantially  larger than in the
spin-0 case and there exists a reasonable sensitivity to  the choice of $\kappa
$. At the LHC with 100 fb$^{-1}$ the search reach for  scalar(S)/vector(V) LQ's
is 1.4/2.2(1.8) TeV for $\kappa=1(0)$. At the Tevatron with 200(2000)
pb$^{-1}$,
the V-LQ reaches are 300(385) GeV for  $\kappa=1$ and 250(330) GeV for
$\kappa=0$. The corresponding results  for S-LQ limits are  170(250). At a 4
TeV
$p\bar p$ collider with 1 fb$^{-1}$ the V-LQ reach is  850 GeV with $\kappa=1$,
while for scalars it is 620 GeV. Correspondingly,  at a 100 TeV $pp$ collider
with 100 fb$^{-1}$ the V-LQ limit is 8.2 TeV with  $\kappa=1$, while for S-LQ's
it is 5.0 TeV. All these results assume a  branching fraction of unity for the
$ej$ final state.

\newpage

Single LQ production has the advantage of a larger available phase space but
has the disadvantage associated with the fact that cross sections are
proportional to an unknown Yukawa coupling, which is expected to be order
electroweak strength or less. In the V-LQ case, the arbitrariness of $\kappa$
still enters the calculation and care must be taken to distinguish possible
$gu$ from $gd$ production. (For example, the $E_6$ type LQ with $Q=-1/3$
is produced via $gu\to LQ+e^+$ and $gd\to LQ+\bar \nu$ with different
Yukawa couplings.) Figs. 20 and 21 show the single LQ rates for both the LHC
and
Tevatron for either production process; $\kappa=0,1$ are considered in the
V-LQ case. It is assumed that all Yukawa couplings are exactly equal to
$\alpha$ in these figures. Clearly, if the Yukawa's are significantly smaller
pair production will generally win out over single production for the entire
mass range.

\vspace*{-0.5cm}
\nn
\begin{figure}[htbp]
\centerline{
\psfig{figure=vlq3a.ps,height=10cm,width=8cm,angle=90}
\hspace*{-5mm}
\psfig{figure=vlq3b.ps,height=10cm,width=8cm,angle=90}}
\vspace*{-1cm}
\caption{\small Single (a)S-LQ (b)V-LQ production at the Tevatron for unit
scaled Yukawa coupling. In (a) the dotted(dashdotted) curve is for $gu(gd)$
production. In (b), the upper(lower) pair of curves corresponds to $gu(gd)$
production with $\kappa=1,0$ respectively.}
\end{figure}

\vspace*{-0.5cm}
\nn
\begin{figure}[htbp]
\centerline{
\psfig{figure=vlq4a.ps,height=10cm,width=8cm,angle=90}
\hspace*{-5mm}
\psfig{figure=vlq4b.ps,height=10cm,width=8cm,angle=90}}
\vspace*{-1cm}
\caption{\small Same as the previous figure but for the LHC.}
\end{figure}

\vspace*{0.4mm}

There are other mechanisms that produce LQ's at hadron colliders.  In an
alternative supersymmetric version of the  LRM {\cite {alrm}}, which takes
advantage of a well known ambiguity in the fermion quantum number assignments
within the {\bf 27}  representation of $E_6$, $W_R$'s can only be produced in
association  with leptoquarks(LQ) at hadron colliders. The basic  process is
$gu\to W_R+LQ$. The discovery limit of order  1.2-2.5 TeV is obtained in the
background-free case at the LHC for an integrated  luminosity of 100 fb$^{-1}$.

\newpage

A detailed study has been performed for this report {\cite {chia}},  in order
to
ascertain the extent to which the backgrounds from top and SM $W$ boson
leptonic
decays masked the signal at the LHC. It was assumed that the  final state
neutral lepton appears as missing energy and that the signals and  backgrounds
are well modeled via PYTHIA.

If one completely ignores the associated LQ,  the signal corresponds to an
increase in the overall lepton  $p_t$ distribution in the region corresponding
to the $W_R$'s Jacobian peak.  For $p_t>M_{W_R}$, the signal surpasses the
backgrounds from $t\bar t$ production  with rates of order 10-100 events  a
year
assuming $M_{W_R}$=1 TeV and $\Delta p_t$=100 GeV.  However, this signal is
overwhelmed by the inclusive lepton spectrum from the  decay of the SM $W$
boson
(due to the fact that at large $p_t$ associated $W_L+j$  dominates) by an order
of magnitude.

This situation was found to be somewhat improved if the leptonic decay modes
of the LQ were included, \ie, $LQ \to ue+\chi^0,~de+\chi^+$, where
$\chi^+(\chi^0)$ is a chargino (neutralino or LSP). The signal now corresponds
to a charged lepton pair plus a jet plus missing energy. Unfortunately,
although the background from $W_L$ is now removed, that from $t\bar t$ still
remains and swamps the signal by two orders of magnitude for a 1 TeV $W_R$
and typical SUSY partner masses.

The conclusion of the analysis is that this  final state is not suitable for
$W_R$ discovery for small LQ masses  (\ie, below those of the SUSY particles)
due to the tiny leptonic  branching fraction possessed by the LQ in this case.

\newpage

\subsection{Single Dilepton Production at $e^+e^-$ Colliders}

Dileptons arise in theories where the gauge group for leptons is expanded
from the SU(2)$_{\sss L}$ of the SM to SU(3). They can appear
as both scalars and as vector gauge particles, and can be singly- or
doubly-charged. The production of doubly-charged scalar and vector dileptons
in $e^+e^-$, $\gamma\gamma$ and $e\gamma$ colliders is summarized for this
report {\cite {lepo}}; for original references, see~{\cite {Framp2,orig}}.

Dilepton interactions are described by the Lagrangian
\begin{eqnarray}
{\cal L}_{int} & = & -\frac{g_{3l}}{\sqrt{2}} \, X_\mu^{++} \, e^\smallT C
\gamma^\mu \gamma_5 e + \frac{g}{\sqrt{2}} \, X^{++} \, e^\smallT
C(1-\lambda \gamma _5)e + h.c.
\end{eqnarray}
where $X_\mu^{++}$ ($X^{++}$) is the vector (scalar) dilepton field, and
$C$ is the charge conjugation matrix. The vector coupling of the vector
dilepton vanishes by Fermi statistics. The coupling of the scalar dilepton
is chiral, so that $\lambda = \pm 1$. The coupling constants $g$ and
$g_{3l}$ are considered as free parameters.

The cross sections for the processes $e^-\gamma
\to X^{--}_{\sss S,V} e^+$, using $g_{3l}^2 \equiv 4\pi \kv \alpha_{em}$ and
$g^2 \equiv 4\pi \ks \alpha_{em}$ is given by ($\alpha$ and $\beta$ are given
in eq.~30).
\begin{eqnarray}
\sigma_{\sss S}(s) & = & \frac{\pi \ks \alpha^2_{em}}{s} \left[ \beta
\left( \frac{3}{2} + \frac{17}{2}{\mx^2\over s} \right) + 8{\mx^2\over s}
\, \ln \left(\frac{2 - \alpha - \beta}{2 - \alpha + \beta}
\right) \right. \nonumber \\
& & \hspace{2cm} \left. + \; \frac{s^2 - 2 s \mx^2 +2 \mx^4}{s^2}
\, \ln \left(\frac{\alpha + \beta}{\alpha - \beta} \right) \right] \,
\\
\sigma_\smallV(s) & = & \frac{\pi \kv \alpha^2_{em}}{s} \left[ \beta
\left(2 + \frac{8s}{\mx^2} + \frac{13}{2}\frac{\mx^2}{s} \right) +
\frac{s^2 - 2s\mx^2 + 2\mx^4}{s^2}
\, \ln \left(\frac{\alpha + \beta}{\alpha - \beta} \right)
\right. \nonumber \\
& & \hspace{2cm} \left. + \; \frac{(-s^3 + 12s^2\mx^2 + 18s\mx^4 -
4 \mx^6)}{2s^2 \mx^2}
\, \ln \left(\frac{2 -\alpha - \beta}{2 - \alpha + \beta} \right) \right]
\nonumber
\end{eqnarray}
%

The explicit electron mass regulates the logarithmic collinear divergence
occurring in that region of phase space in which all particles go down
the beam pipe. (The apparent divergence in the massless electron limit can
be removed by detector cuts since these will be present in any real cross
section determination.) Thus, the bulk of the above cross sections are due to
those
events in which the only particles detected are the decay products of the
$X^{--}_{\sss S,V}$ ($e^-e^-$, $\mu^-\mu^-$ or $\tau^-\tau^-$). This
results in an unmistakable signature with virtually no SM background.

For the process $e^+e^- \to X^{--}_{\sss S,V} e^+e^+$, an energetic virtual
photon is emitted from the $e^+$ beam, which then becomes part of an
$e^-\gamma$ collision. To calculate this cross section, we use a photon
distribution function for the virtual photon, and fold it together with the
previously determined cross section of $e^-\gamma \to X^{--}_{\sss S,V}
e^+$, using numerical integration.
The process $\gamma\gamma \to X^{--}_{\sss S,V} e^+e^+$ proceeds along
similar lines as above, except that a virtual electron is emitted from the
$\gamma$ beam, and a fermion distribution function is used.

In this summary, we show the production cross sections for vector and
scalar dileptons in each process, at 1 TeV.
Assuming a luminosity of 60 fb$^{-1}$, we suppose that 25
events, i.e.\ a cross section of 0.4 fb, are required for discovery. We
take the coupling constants, $g_{3l}$ and $g$, to be of electromagnetic
strength, i.e.\ $\kv = 1$ and $\ks = 1$. Of course, since the cross
sections are linear in $\kv$ and $\ks$, these graphs can easily be scaled
to other parametric values of the couplings.

In Fig. 22a we see that the cross section for scalar dilepton production in
$e\gamma$ collisions is huge, orders of magnitude above the discovery
limit. Thus, scalar dileptons with masses virtually up to the kinematic
limit are observable, even for couplings as small as $\ks = 5$-$7 \times
10^{-4}$. The cross sections in $e^+e^-$ and $\gamma\gamma$ collisions are
smaller, but still large enough for the observation of dileptons with
masses approaching the kinematic limit.
Fig. 22b shows the production cross sections for vector dileptons. Once
again, the cross section in the $e\gamma$ process is clearly much larger
than that of the other two processes, so that dileptons of masses up to the
kinematic limit and couplings as small as $\kv = 3$-$4 \times 10^{-4}$ can
be detected. Dileptons can be observed in $e^+e^-$ and $\gamma\gamma$
collisions for $\kv$ as low as $\sim 0.01$.

This analysis shows that both scalar and vector dileptons can be
easily observed in all three modes of the $e^+e^-$ collider. The bulk of
the cross sections comes from those events in which the only particles
detected are the two leptons coming from the decay of the dilepton, an
unmistakable signature.

\nn
\vspace*{-3mm}
\hspace*{-0.5cm}
\begin{figure}[htbp]
\centerline{
\psfig{figure=s_summary1000.ps,height=3.5in,width=2.8in}
\hspace*{0.2in}
\psfig{figure=v_summary1000.ps,height=3.5in,width=2.8in}}
\vspace*{-0.5cm}
\caption{\small Production cross section for (a) scalar (left) and  (b) vector
(right) dileptons in the processes $e^-\gamma \to X_{\sss S}^{--}e^+$ (solid
line), $e^-e^+ \to X_{\sss S}^{--}e^+e^+$ (dashed line) and $\gamma\gamma \to
X_{\sss S}^{--} e^+e^+$ (dash-dot line), for a 1 TeV NLC with $\ks = 1$. The
horizontal line is the assumed discovery cross section of 0.4 fb.}
\end{figure}
\vspace*{0.5cm}

\section{New Interactions}

\subsection{Top Quark Anomalous Chromomagnetic Moments}

The discovery of the top quark at the Tevatron~{\cite {cdf1}} by the
CDF and D0 Collaborations has renewed  thinking about what may be learned from
a
detailed study of the properties of this particle.  One point of view is that
this discovery represents a great triumph and  confirmation of the predictions
of the SM,  in that the top quark lies in the mass range anticipated by
precision electroweak data~{\cite {pew}}. Another viewpoint is that  the
subtleties of top quark physics itself may shed some light on new  physics
beyond the Standard Model.

Amongst others, one set of the top quark properties which deserve study are
its
couplings to the various gauge bosons;  up until recently such analyses~{\cite
{old}} have concentrated on the  electroweak couplings of the top, \ie, ~its
interactions with the $W$, $Z$ and  $\gamma$. In what follows, we consider the
possible  existence of an anomalous chromomagnetic moment, a dimension-5
coupling,  $\kappa$, at the  $t\bar tg$ vertex and we explore the capability of
the Tevatron, LHC  and NLC to probe  this coupling. Such anomalous interactions
may arise with a  reasonable strength in extended technicolor  or compositeness
scenarios~{\cite {ak}} and, \eg,~may lead to significant  alterations in the
top
production cross section at the Tevatron and other  colliders. In such
scenarios, the chromomagnetic moment is usually induced as  a natural
by-product
of the top quark mass generation process.  At the present time the CDF and D0
top cross section results seem to be in rough agreement with, although still
somewhat larger than, the expectations
of QCD~{\cite {laenen}}. The original version of this analysis was motivated
by
this somewhat larger than expected result first obtained by CDF last year.
Previous to the
present analysis, only rather weak limits on $\kappa$ (of order 10) existed, in
particular, from operator mixing contributions to the  $b\to s\gamma$ decay;
see the last paper in~{\cite {old}}).
Data from the  Tevatron, LHC, and NLC
will be able to improve this sensitivity by two orders  of magnitude. For
details of the  analyses presented below, see ref.~{\cite {tomtop}}.

The piece of the Lagrangian which governs the $t\bar t g$ and
$t\bar t gg$ couplings is:
\begin{equation}
{\cal L}=g_S\bar t T_a \left( \gamma_\mu G_a^\mu+i{\kappa\over
{2m_t}}\sigma_{\mu\nu}G_a^{\mu\nu} \right)t  \,,
\end{equation}
where $g_S$ and $T_a$ are the usual $SU(3)_C$ coupling and
generators, $m_t$
is the top quark mass, $G_a^\mu(G_a^{\mu\nu})$ is the gluon field (strength
tensor), and $\kappa$ is the anomalous chromomagnetic moment which is zero in
the SM. Note that $SU(3)_C$ gauge invariance requires that {\it both} vertices
be modified when $\kappa$ is present.

Turning first to the examination of  the effects of non-zero $\kappa$ on
$t \bar t$ production at hadron colliders, we present
the parton-level $q\bar q \to t\bar t$ and $gg \to t\bar t$
differential cross sections. (For single top production at these machines, we
need the corresponding $gW\to t\bar b$ result.) We note in passing that the
$q\bar q$ process dominates (about $90\%$) at the Tevatron while the $gg$
one is similarly dominant at the LHC. This analysis has shown that the total
top cross section is the most sensitive quantity in probing $\kappa$
at hadron colliders provided it is not too large.
For $q\bar q \to t\bar t$, which occurs through $s$-channel gluon
exchange, one obtains
\begin{small}
\begin{equation}
{d\sigma_{q\bar q}\over {d\hat t}}={2\pi\alpha_S^2\over {27\hat
s^2}}\left[
\left(1+{2m_t^2\over {\hat s}}\right)+3\kappa+\kappa^2\left({\hat s\over
{8m_t^2}}
+1\right)+{1\over {4}}(3z^2-1)\left(1-{\hat s\over
{4m_t^2}}\kappa^2\right)
\right]  \,,
\end{equation}
\end{small}
with $\hat s$ being the parton level c.m. energy and $z$ the
cosine of the corresponding scattering angle.
The case of the $gg \to t\bar t$ is
more complicated since it proceeds through $s$-, $t$-, and $u$-channel diagrams
as well as a contact term which is present due to gauge
invariance; defining the kinematic abbreviations $x = m_t^2/ {\hat s},
K = \kappa/(2\sqrt{x})$ and $ d = 1-z^2+4xz^2$, the resulting differential
cross section can be written as
\begin{equation}
{d\sigma_{gg}\over {d\hat t}}={\pi\alpha_s^2\over {64\hat
s^2}}\left[
T_0+T_1K+T_2K^2+T_3K^3+T_4K^4\right]  \,,
\end{equation}
which is a quartic polynomial in $\kappa$, where the $T_i$ coefficients
can be written as
\begin{eqnarray}
T_0 &=& 4(36xz^2-7-9z^2)(z^4-8xz^4+16x^2z^4-32x^2z^2+8xz^2-8x-1)
/{3d^2} \,, \nonumber \\
T_1 &=& -32(36xz^2-7-9z^2)\sqrt {x}/{3d} \,, \nonumber \\
T_2 &=& -16(72x^2z^2-46xz^2+7z^2-16x-7)/{3d} \,, \\
T_3 &=& 32(-7z^2+28xz^2-5x+7)\sqrt{x}/{3d} \,, \nonumber \\
T_4 &=& 16(-8xz^4+16x^2z^4+z^4-4x^2z^2+9xz^2-2z^2+1-x+4x^2)/{3d}
\,. \nonumber
\end{eqnarray}

\vspace*{-0.5cm}
\nn
\begin{figure}[htbp]
\centerline{
\psfig{figure=toptev.ps,height=7cm,width=8cm,angle=90}
\hspace*{-5mm}
\psfig{figure=toplhc.ps,height=7cm,width=8cm,angle=90}}
\vspace*{-1cm}
\caption{\small (a) NLO cross sections for the $q\bar q \to t\bar t$
(dash-dotted) and $gg \to t\bar t$ (dotted) subprocesses as well as the total
cross section (solid) at the Tevatron as functions of $\kappa$ for $m_t=170$
GeV using the CTEQ parton distribution functions. The horizontal dashed lines
provide the $\pm 1\sigma$ CDF cross section determination while the horizontal
dotted line is the D0 $95\%$ CL upper limit. (b) Same as (a) but for the LHC
and with the roles of $gg$ and $q\bar q$ interchanged.}
\end{figure}

\vspace*{-0.5cm}
\nn
\begin{figure}[htbp]
\centerline{
\psfig{figure=gwtbtev.ps,height=7cm,width=8cm,angle=90}
\hspace*{-5mm}
\psfig{figure=gwtblhc.ps,height=7cm,width=8cm,angle=90}}
\vspace*{-1cm}
\caption{\small Single top cross section at the (a) Tevatron and (b) LHC. From
top to bottom, the curves correspond to $\kappa=2,-2,1,-1,0$.}
\end{figure}

As in the $q\bar q$ case, the $gg \to t\bar t$ cross section increases as
$\kappa$ increases in the positive direction. These cross sections the for
Tevatron and LHC are  shown in Fig. 23a-b, respectively. If the SM cross
section
is realized, accounting for various uncertainties (parton densities, NNLO
terms, scales, luminosity, statistics, etc.) the $95\%$ CL ranges for $\kappa$
assuming Tevatron luminosities of ${\cal L}=100(250,500,1000)$ pb$^{-1}$  are
-0.14 to 0.15, -0.11 to 0.12,  -0.09 to 0.11, and -0.08 to 0.11, respectively,
\ie, they are systematics  limited  at large  luminosity. The Tevatron analysis
was then extended to the LHC case where it  was found that the results were
clearly systematics limited at the level of  $\kappa \simeq \pm 0.10$ due to
the
uncertainties from higher order QCD  corrections and parton density variations.

In the case of single production of top, we expect low sensitivity to
$\kappa$ due to the dominance of the light $b$-quark exchange diagram. This
is clearly the case as shown in Figs.~24a-b for the Tevatron and LHC.

At the NLC, the $t\bar tg$ vertex can only be directly explored via the QCD
radiative process $e^+e^- \to t\bar tg$. Relative to the LHC and Tevatron,
this
results in a substantial loss in statistics which can be compensated for  by
the
cleanliness of the environment as well as a reduction in the associated
theoretical uncertainties. Since the new  $\kappa$-dependent interaction is
proportional to the gluon 4-momentum, one is  lead to a study of the gluon
energy distribution associated with  $t\bar t$ production.

To leading order in $\alpha_S$ one can factorize this cross section into
separate contributions due to the vector and axial-vector couplings of the top
quark to the $s$-channel exchanged gauge bosons as
\begin{eqnarray}
d^2W/ dz_1dz_2 = F_v \ d^2W_v/ dz_1dz_2 +F_a\ d^2W_a/ dz_1dz_2 \,,
\end{eqnarray}
where $F_{v,a}$ are the `weighting' factors.
This result is scaled to the lowest order $t\bar t$
production cross section, \ie, $W=\sigma/\sigma_0$, where
$\sigma_0=\sigma(e^+e^- \to t\bar t)$, and

\begin{eqnarray}
F_v = {{1\over {2}}\beta(3-\beta^2)A_v\over {\beta^3 A_a+
{1\over {2}}\beta(3-\beta^2)A_v}} \ \, , \ \
F_a &=& {\beta^3 A_a\over {\beta^3 A_a+{1\over {2}}\beta(3-\beta^2)A_v}} \,,
\end{eqnarray}
with $\beta=\sqrt {(1-4m_t^2/s)}$, and
\begin{eqnarray}
A_v = \sum_{ij} ~(v_iv_j+a_ia_j)_e(v_iv_j)_t P_{ij} \,, \ \
A_a = \sum_{ij} ~(v_iv_j+a_ia_j)_e(a_ia_j)_t P_{ij} \,, \nonumber \\
P_{ij} = s^2 {[(s-M_i^2)(s-M_j^2)+(\Gamma M)_i(\Gamma M)_j]
\over {[(s-M_i^2)^2+(\Gamma M)_i^2][(s-M_j^2)^2+(\Gamma M)_j^2]}} \,.
\hspace*{2cm}
\end{eqnarray}
The sum in the expression above is over the $s$-channel $\gamma~(i,j=1)$ and
$Z~(i,j=2)$ gauge boson exchanges. Defining the overall normalization
coefficients
\begin{small}
\begin{eqnarray}
N_v = {2\alpha_s\over {3\pi}}(2m_t^2s^2x_1^2x_2^2)^{-1}[{1\over{2}}\beta
(3-\beta^2)]^{-1} \,, \ \
N_a = {2\alpha_s\over {3\pi}}(2m_t^2s^2x_1^2x_2^2)^{-1}[\beta^3]^{-1} \,,
\end{eqnarray}
\end{small}
where $x_i=1-z_i$, we obtain the complete $t\bar tg$ double differential
cross section:
\begin{eqnarray}
{d^2W_v\over {dz_1dz_2}} &=& N_v\left[ -8m_t^6(x_1+x_2)^2-4sm_t^4
[x_1^2(1+2x_2)+x_2^2(1+2x_1)] \right. \nonumber \\
 & & + 2s^2m_t^2x_1x_2 [(1-x_1)^2+(1-x_2)^2+\kappa(x_1-x_2)^2] \nonumber \\
 & & \left. +\kappa^2s^3x_1^2x_2^2
(1-x_1-x_2)\right] \,, \nonumber \\
{d^2W_a\over {dz_1dz_2}} &=& N_a\left[ 16m_t^6(x_1+x_2)^2+2sm_t^4
[(\kappa^2+2\kappa+2)x_1x_2(x_1+x_2)^2 \right.  \nonumber \\
 & & + 8x_1x_2 (x_1+x_2)-2(x_1^2+x_2^2+6x_1x_2)]+2m_t^2s^2x_1x_2[(1-x_1)^2
 \nonumber \\
 & & + (1-x_2)^2+\kappa(x_1^2+x_2^2-4)+\kappa^2x_1x_2(x_1+x_2-3)] \nonumber \\
 & & \left. + \kappa^2s^3x_1^2x_2^2(1-x_1)(1-x_2)\right] \,
\end{eqnarray}
The dominant effect of $\kappa \neq 0$ is to induce an increase in
the high energy tail of this distribution.
This same energy dependence leads
to the observation that the finite $\kappa$ contributions grow rapidly with
increasing $\sqrt {s}/2m_t$, implying increased sensitivity at an NLC with
$\sqrt {s}=1$ TeV instead of 500 GeV. In this first study, we ignore
effects from top decay (except in the statistics) and perform a LO analysis.
Estimates of contributions from higher order are lumped into the
uncertainties when obtaining limits. Fig. 25 shows this distribution for
the case of $\sqrt {s}=1$ TeV for $\alpha_S=0.10$ while Fig. 26a shows the
result of integrating this distribution for values of
$z=2E_{gluon}/{\sqrt {s}}>0.4$. Assuming that the SM results are realized,
bounds on $\kappa$ may be
obtainable by either ($i$) counting excess events with high energy gluon jets
or ($ii$) by a fit to the gluon energy distribution via a Monte Carlo analysis.
Events are selected with at least one b-tag as well as one high $p_t$ lepton
and gluon jet energies larger than 200 GeV. Such large jet energies will allow
a clean separation from the top decays and will simultaneously place us in
the region of greatest $\kappa$ sensitivity. For a luminosity of 200 fb$^{-1}$
the resulting $95\%$ CL allowed range is found to be $-1.0\leq \kappa \leq
0.25$. Substantial improvement is obtained by fitting the spectrum itself;
Fig. 26b shows the Monte Carlo generated spectrum and best fit($\kappa=0.06$)
assuming that the SM is realized. At $95\%$ CL, one now obtains the allowed
range of $-0.12\leq \kappa \leq 0.21$ for the same luminosity as above.

The LHC, Tevatron and  NLC provide complementary windows on the possible
anomalous chromomagnetic couplings of the top with different systematics.

\nn
\vspace*{-5mm}
\hspace*{-0.5cm}
\begin{figure}[htbp]
\centerline{\psfig{figure=y1.ps,height=8cm,width=12cm,angle=90}}
\vspace*{-1cm}
\caption{\small Gluon jet energy spectrum assuming $\alpha_s=0.10$ for
$m_t=175$ GeV at a 1 TeV NLC. The upper(lower) dotted, dashed, and
dot-dashed curves correspond to $\kappa$ values of 3(-3), 2(-2), and 1(-1)
respectively while the solid curve is conventional QCD with $\kappa=0$. }
\end{figure}
\vspace*{-0.3cm}
\nn
\begin{figure}[htbp]
\centerline{
\psfig{figure=y2a.ps,height=6cm,width=8cm,angle=90}
\hspace*{-5mm}
\psfig{figure=y2b.ps,height=6cm,width=8cm,angle=90}}
\vspace*{-1cm}
\caption{\small
(a)Integrated gluon energy spectrum for the same input parameters and labeling
as in the previous figure as  a function of $\kappa$ assuming $z_{cut}=0.4$.
(b)Best fit gluon spectrum through the points generated by the Monte
Carlo analysis for $\kappa=0.06$.}
\end{figure}

\subsection{Top quark radius and anomalous magnetic moment}

One of the most natural consequences of substructure in the fermionic  sector
is
the appearance of an intrinsic finite size of quarks and leptons  due to the
interaction of the $preons$, the more elementary constituents \cite{X1}.  The
search for non--zero radii and anomalous magnetic moments has been pursuit
since
a long time and stringent bounds on the	size of	electrons and muons have been
set by measurements of	$(g-2)_{e,\mu}$	\cite{X2}. Any non point--like structure
of these particles is	restricted to energy scales above 1 TeV. Similar
limits can be obtained for light quarks	from analyses of quark--quark
scattering
in $p\bar{p}$ colliders	\cite{X3} and for $\tau$ leptons	and $b$	quarks
from the	high--precision	LEP measurements \cite{X4}.

These bounds for the light fermions cannot be readily extrapolated to the heavy
top quark. Indeed, as	its large mass seems to	indicate, the top quark could
play a special r\^ole and it may be the	first place where non--standard effects
will appear. (In models where the standard gauge symmetry is dynamically broken
by	$\bar{t}t$ condensates, anomalous couplings are	also expected to occur
\cite{X5}.) Here, we  discuss the
potential of a 0.5 TeV $e^+e^-$ collider in probing
anomalous couplings of the  top quark to electroweak gauge bosons.

The general coupling of a gauge boson $\alpha=\gamma,Z$ to fermions can	be
written as
\begin{eqnarray}
ie_0 \left[ f_1^\alpha \gamma_\mu +\frac{i}{2m_f}f_2^\alpha
\sigma_{\mu\nu}q^\nu+  f_3^\alpha \gamma_\mu \gamma_5 +
\frac{i}{2m_f}f_4^\alpha \sigma_{\mu\nu}\gamma_5 q^\nu\right]
\end{eqnarray}
\noindent with $e_0$ being the proton charge and $q$ the momentum
carried	by the gauge boson. In principle, there	are also $q^\mu$ and
$q^\mu \gamma_5$ couplings, but	they give vanishing contributions if
the gauge boson	is on shell or couples to massless fermions. In the
point--like limit, the form factors $f_1$--$f_4$ reduce	to the usual
standard model couplings: the CP violating terms $f_4^{\gamma,Z}$ are
absent and at the tree level $f_2^{\gamma,Z}$ are equal	to zero, while
$f_1^\gamma = e_f, f_3^\gamma=0$ and $f_1^Z =v_f^Z, f_3^Z = a_f^Z$.
The form factors $f_{1,3}$ are	related	to the radius $R$ which
is proportional	to the compositeness scale $R=\sqrt{6/\Lambda}$: $f_{1,3
}^\alpha \sim (f_{1,3}^\alpha)^{SM}(1+s/\Lambda^2)$ ; they could in
principle be different for $\gamma$ and	$Z$ couplings. Note that the
radius used here is the	physical particle radius which is not plagued
by the ambiguities due to the unknown coupling constants as it is the
case in	contact	terms. $f_2^{\gamma,Z}$	are the	anomalous magnetic
moments	which, in chiral theories  as suggested	by $(g-2)_{e,\mu}$,
are proportional to $m_f^2/\Lambda^2$.

With the general form given above, one can write the most general expression
for the differential cross section d$\sigma$/d$\cos \theta$ in the processes
$e^+ e^- \rightarrow f\bar{f}$ and $\gamma \gamma \rightarrow f\bar{f}$
\cite{X6}.
In $e^+ e^-$ annihilation, this cross section allows three independent
measurements and
one may	choose $R_f$ the total cross section normalized	to $\sigma_0$,
the forward--backward asymmetry	$A_f$ of the fermion and the parameter
$\alpha_f$ defined by d$\sigma$/d$\cos\theta \propto 1+\alpha_f
\cos^2 \theta$; the expressions are given in \cite{X6}.
These	observables have to be compared	to  the	experimental
data once the radiative	corrections of the SM have been	properly
taken into  account. At	a 0.5 TeV $e^+e^-$ collider and	for $m_t=
175$ GeV, we show in Fig.~27 the deviations of the three observables
from the SM expectations as a function of the compositeness scale
$\Lambda$. We have set the CP violating	terms $f_4$ to zero.

\vspace*{-0.5cm}
\begin{figure}[htbp]
\hspace*{-1.2cm}
\centerline{\psfig{figure=ftop1.ps,height=15.5cm,width=13.5cm}}
\vspace*{-1.1cm}
\caption{\small Deviations of the normalized cross section $R_t$, the
forward-backward asymmetry $A_t$ and the $\alpha_t$ parameter from SM
expectations in $\ee \ra t\bar{t}$ at $\protect \sqrt{s}=0.5$ TeV, for
different values of the form-factors $f_i$.}
\end{figure}

Fig.~27a shows the deviations in the case where $\delta f_{1,3}^Z =
\delta f_1^\gamma =s/\Lambda^2$ and $f_2^Z=f_2^\gamma =m_t^2/\Lambda^2$. The
shifts can be very large especially for relatively small values of
$\Lambda$. The ratio $R_t$ is the most sensitive quantity and an experimental
accuracy of 2\% in its measurement allows to probe values of  $\Lambda$
slightly larger than 5 TeV. The forward--backward asymmetry as well as the
$\alpha_t$ parameter are less sensitive to this choice, as shown in
Fig.~27b where we have set $\delta f_1=\delta f_3=0$.	Indeed,	they are much
more sensitive to the anomalous magnetic moments	than to	the change in
the $f_1$ and $f_3$ form factors contrary to $R_t$. In Fig.~27c the same
three observables are shown when  the anomalous couplings to the photon
are switched off (this happens for instance in dynamical symmetry breaking
models where only the couplings  to the $Z$ boson are expected to be
non--universal).  The deviations are much smaller in $R_t$ than in the previous
case (this is due to the fact that the photon exchange dominates	in the
cross section) but they are much larger	in $A_t$. Scales of the	order of 2 TeV
can be  probed in this	case. \s

The possibility of turning the $e^+ e^-$ collider intto a
$\gamma	\gamma$	collider using back--scattered laser beams, can	be
exploited to measure the $\gamma \bar{f}f$ couplings independently
of the $Z \bar{f} f$ couplings.
Assuming the total energy of the $\gamma \gamma$ collider to
be 0.4 TeV, we show in Fig.~28 the deviations in
the total cross	section
as a function of $\Lambda$ in the three	cases: $\delta f_1 =s/\Lambda^2,
f_2=m_t^2/\Lambda^2$; $\delta f_1 =s/\Lambda^2,	f_2=0$ and $\delta f_1
=0, f_2=m_t^2/\Lambda^2$. As it	can be seen, they are larger than in
the $e^+ e^-$ case and the sensitivity to the anomalous	magnetic
moments	is slightly better.

\vspace*{-0.01cm}
\begin{figure}[htbp]
\hspace*{-1.2cm}
\centerline{\psfig{figure=ftop2.ps,height=8.5cm,width=12cm,angle=-90}}
\vspace*{-0.6cm}
\caption{\small Deviations of the total cross section $\gamma \gamma
\ra t\bar{t}$ from the SM value for different values of the form-factors
$f_i$; at $\protect \sqrt{s}_{\gamma \gamma}$ is fixed to 400 GeV.}
\end{figure}

Hence, a 500 GeV c.m. energy $e^+e^-$ collider
is a unique facility to	probe the static properties of the top quark. A
measurement of the total cross section in the $e^+e^-$ or
$\gamma	\gamma$	modes with an accuracy of $2\%$	allows to probe
compositeness scale values up to 10 TeV. This corresponds to a radius of
the order of $10^{-16}$	cm. Furthermore, for light fermions the	present
limits can be greatly  improved.

\subsection{$q\bar q \gamma\gamma$ Contact Interactions and Diphoton
Production }

Instead of  the direct production of new particles, physics beyond the SM may
first  appear as deviations in observables away from SM expectations, such as
in
the  rates for rare processes or in precision electroweak tests. Another
possibility is that deviations in cross sections of order unity may  be
observed
once sufficiently high energy scales are probed. This kind of new  physics can
generally be parameterized via a finite set of non-renormalizable  contact
interactions, an approach which is quite popular in the  literature {\cite
{contact}}. In fact, limits already exist from a number of  experiments on the
scales associated with contact interactions of various  types {\cite {lims}}.
Here {\cite {tgr2g}}, we explore the capability of both the Tevatron and LHC to
probe the existence of flavor-independent (apart  from electric charge),    $q
\bar q \gamma\gamma$ contact interactions of dimension-8. Searches  for such
operators, with the  quarks replaced by electrons, have already been performed
at TRISTAN and LEP {\cite {aleph}} and have resulted in a lower bound of
approximately 140 GeV  on the associated mass scale.

To be definitive, we will follow the notation employed by {\cite {con}} as
well as by the ALEPH Collaboration {\cite {aleph}} and assume that these
new interactions are parity conserving. In this case we can parameterize the
$q \bar q \gamma\gamma$ contact interaction as
\begin{equation}
{\cal L}= 2ie^2 Q_q^2 \Lambda^{-4} F^{\mu\sigma}F_\sigma^\nu \bar q
\gamma_\mu\partial_\nu q  \,,
\end{equation}
where $Q_q$ is the quark charge and $\Lambda$ is the associated mass scale.
The most obvious manifestation of this
new operator is to modify the conventional Born-level partonic $q \bar q \to
\gamma\gamma$ differential cross section so that it now takes the form
\begin{equation}
{d\hat \sigma\over {dz}}=Q_q^4 {2\pi \alpha^2\over {3\hat s}}
\left[{1+z^2\over {1-z^2}} \pm 2{\hat s^2\over {4\Lambda_{\pm}^4}}(1+z^2)+
\left({\hat s^2\over {4\Lambda_{\pm}^4}}\right)^2 (1-z^4)\right]  \,,
\end{equation}
where $\hat s,~z$ are the partonic c.m. energy and cosine of the
c.m. scattering angle, $\theta^*$, respectively. Note that we
have written $\Lambda_{\pm}$ in place of $\Lambda$ in the equation above
to indicate that the limits we obtain below will depend upon whether
the new operator constructively or destructively interferes with the SM
contribution.

There are two major
effects due to finite $\Lambda$: ($i$) Clearly, once $\hat s$
becomes comparable to $\Lambda^2$, the parton-level differential
cross section becomes less
peaked in the forward and backward directions implying that the photon pair
will generally be more central and will occur with higher average $p_t$'s.
($ii$) When integrated over
parton distributions the resulting cross section will lead to an increased
rate for photon pairs with large $\gamma\gamma$ invariant masses. Thus we
employ strict $\eta$ and $p_t$ cuts on both photons to reduce backgrounds
from SM processes.

In presenting numerical results,
we integrate the invariant diphoton mass distribution above a given fixed
minimum value of the diphoton mass, $M_{\gamma\gamma}^{min}$, subsequent to
making all the other cuts. In order to get an estimate for the
event rates involved, we scale this integrated cross section by a luminosity
appropriate to the Tevatron or the LHC, \ie, 20 pb$^{-1}$ and 100 fb$^{-1}$,
respectively.
Figs. 29a-b compare the SM diphoton cross section as a function of
$M_{\gamma\gamma}^{min}$ with the constructive interference scenario for
various values of the $\Lambda$ parameter.

Assuming that no event excesses are observed, we can ask
for the limits that can be placed on $\Lambda_{\pm}$ as the Tevatron
integrated luminosity is increased. To do this we perform a Monte Carlo
study, first dividing the $M_{\gamma\gamma}^{min}$ range above 100 GeV into
nine steps of 50 GeV. Events are generated
using the SM as input and are then fit to the resulting
$\Lambda_{\pm}$ dependent fitting function. For a luminosity
of $ 100 (250, 500, 1000, 2000)$ pb$^{-1}$ we obtain the
bounds
$\Lambda_+ >  487 (535, 575, 622, 671)$ GeV
and $\Lambda_- > 384 (465, 520, 577, 635)$ GeV, respectively, at
$95 \%$ CL.
Correspondingly, for a similar analysis at the LHC, we find with an integrated
luminosity of 100 fb$^{-1}$, the $95\%$ CL
bounds of $\Lambda_+>2.83$ TeV and $\Lambda_->2.88$ TeV.

\vspace*{-0.6cm}
\nn
\begin{figure}[htbp]
\centerline{
\psfig{figure=dipho1.ps,height=7cm,width=8cm,angle=90}
\hspace*{-6mm}
\psfig{figure=dipho2.ps,height=7cm,width=8cm,angle=90}}
\vspace*{-0.6cm}
\caption{\small (a) Diphoton pair event rate, scaled to a luminosity of 20
pb$^{-1}$, as a function of $M_{\gamma\gamma}^{min}$ at the
Tevatron subject to the cuts discussed in the text. The solid curve is the
QCD prediction, while from top to bottom the dash dotted curves correspond to
constructive interference with the SM and
a compositeness scale associated with the $q\bar q \gamma\gamma$ operator of
$\Lambda_+=0.2,~0.3,~0.4,~0.5,$ and 0.6 TeV respectively.
(b) Same as (a), but for the LHC scaling to a lumonosity of 100
fb$^{-1}$. From top to bottom the
dash dotted curves now correspond to $\Lambda_+=0.75,~1.0,~1.25,~1.5,~1.75$ and
2.0 TeV respectively.}
\end{figure}

\vspace*{-4mm}

\subsection{New resonant structures}

\def\mz{M_Z^2}
\def\mw{M_W^2}
\def\sef{s^2_{EFF}(\mz) }
\def\eps{\epsilon}
\def\q{q^2}
\def\p{ {\cal P}}
\def\g{\gamma}

In this subsection we will discuss the indirect effects of Technicolour-like
vector particles that are strongly coupled to the SM gauge bosons in the
process $\ee \ra\bar{f}f$. The high--precision LEP1 data have already set
rather stringent bounds on Technicolour models \cite{peskin}, based on their
effect to the quantity $S$ \cite{peskin} a combination of the one--loop
SM vector boson self--energies. Recently, a general formalism has been
established \cite{strong} which allows to calculate the relevant one--loop
self--energy corrections to processes at high energies $e^+e^-$ colliders.

The main idea is that of expressing the various effects in the form of a
once-subtracted dispersion integral, and of fixing the necessary subtraction
constants by suitable model-independent LEP1 results. In this way, one is led
to a compact representation of several observables which presents two main
advantages. The first one is that it allows to express new physics
contributions through convergent integrals. The second one is that LEP1
constraints are automatically incorporated in the expressions of the
observables. For example, the cross section  for muon production at a c.m.
energy $\sqrt{q^2}$, $ \sigma_\mu(\q)$, at the one--loop level takes the form
\begin{eqnarray}
\sigma_\mu^{SE}(\q)&=&
{4\pi\q\over3}\biggm\{  [{\alpha(\mz)\over q^2}]^2[1+2D_\gamma(\q)]
+ {1\over(q^2-\mz)^2+\mz\Gamma^2_Z} \nonumber \\
 & &\null \bigm[{3\Gamma_l\over M_Z}\bigm]^2[ 1-2 D_Z(\q)
-{16 s^2_1v_1\over1-v^2_1}D_{\gamma Z} (q^2)] \biggm\}
\end{eqnarray}
where $\Gamma_l$ is the leptonic $Z$ width, $\alpha(\mz)=[1\pm0.001]/128.87$
and
\begin{eqnarray}
D_\gamma(\q) & =& -(\q-\mz)/\pi
\p\int_0^\infty ds\ Im\ F_\gamma(s) (s-\q)^{-1}(s-\mz)^{-2} \nonumber \\
D_Z(\q) & =&  (\q-\mz)/\pi\ \p\int_0^\infty
ds s Im\ F_{ZZ}(s) (s-\q)^{-1}(s-\mz)^2 \nonumber \\
D_{\gamma Z}(\q) & =& (\q-\mz)/\pi\ \p\int_0^\infty ds\ Im\ F_{\kappa'}(s)
(s-\q)^{-2}(s-\mz)^{-1}
\end{eqnarray}
with $F_\kappa'=c_1/s_1, F_{Z\g}, s^2_1 c^2_1=\pi\alpha/(\sqrt2 G_\mu\mz)$
and $s^2_1=1-c^2_1\simeq 0.217, v_1=1-4s^2_1$.
The imaginary parts which appear in these expressions are  constructed from the
self-energies; for Technicolour models, they are separately gauge-invariant.
Similar representations can be established for several other observables
like forward-backward and polarization asymmetries, $etc..$
For each observable, one finally obtains an
expression that include the full effect of the oblique correction at one-loop
in the form: $O(\q)= c_0[1+c_{\gamma}D_{\gamma}(\q)+c_ZD_Z(\q)+c_{\gamma Z}
D_{\gamma Z}(\q)]$, where the analytic expressions of the various coefficients
can be found in \cite{strong}.

One can use this formalism to calculate the possible effects of a pair of
vector
(V) and axial vector (A) resonances strongly coupled to the photon  and to the
$Z$ boson. The parameters which enter the expressions of the imaginary parts of
the various spectral functions are the couplings $F_{V,A}$ and the masses
$M_{V,A}$  (assumed to be larger than $\sqrt{q^2}$). Two different
theoretical models have been considered:

(I) A Technicolour--like framework in which the validity of the two Weinberg
sum rules \cite{weinberg} are exploited. Only their very general consequence,
i.e. the positivity of $S$ are retained. In a zero--width approximation (in
practice, one needs to use a finite width description of the V,A
resonances) one has: $S= 4\pi[F^2_V/ M^2_V - F^2_A/ M^2_A] = 4\pi (F^2_{\pi}
/M^2_V)[1+M^2_V/ M^2_A]$. The present constraint on $S$ is $-0.9 \leq S \leq
0.4$ \cite{Sexp}. In this model only the  positive upper bound is effective.

(II) The constraints due to the Weinberg sum rules are released, a choice
which has the consequence of introducing one more degree of freedom since it
eliminates the theoretical relation between $F_V$ and $F_A$. As a consequence,
$S$ can now take negative values, and in addition the strength of the ratios
$F_V/M_V$ and $F_A/M_A$ is no more bounded. The limiting case of a strongly
interacting regime for which the value of $F_V/M_V$ is twice the QCD value,
$ F_V / M_V=2 f_\rho / m_\rho =1/ \sqrt{2\pi}$, has been considered. Then,
for every choice of $F^2_V/M^2_V$, $F^2_A/M^2_A$ is allowed to saturate both
limits imposed by the bounds on $S$.

\nn
\begin{figure}[htbp]
\vspace*{-6mm}
\centerline{
\psfig{figure=lay1.ps,height=8.5cm,width=8cm}
\hspace*{-5mm}
\psfig{figure=lay2.ps,height=8.5cm,width=8cm}}
\vspace*{-2.5cm}
\caption{\small (a) Limits on $M_A$ $vs$ $M_V$ at 1 TeV from $\sigma_\mu$(dots)
$A_{LR}$ (dot--dash) and $A_{\tau}$ (dashes) using the Weinberg sum rules and
data on $S$. The lighter shaded region is the result of quadratically combining
the two leptonic limits and the darker region combines all constraints. The two
solid lines correspond to $M_A=(1.1,1.6)M_V$. (b) Limits obtained when released
from the Weinberg sum rules but imposing the $F_V/M_V$ limitation from
$\sigma_\mu$ (vertical,dotted), $A_\tau$ (vertical dashed), $A_{LR}$
(dotdashed), $R_{b,\mu}$ (shortdashed), $R^{(5)}$ (dotted), $A_{FB,\mu}$
(long dashed). The solid lines show $M_A=(1.0,1.6)M_V$.}
\end{figure}

Assuming a certain accuracy  for the measurement of each observable, one
accordingly obtains the observability limit of the self-energy effect that is
translated in an upper bound on the masses $M_{V,A}$ . For a 1 TeV $e^+e^-$
collider the assumed accuracies are of  a relative one percent for
$\sigma_\mu$, $A_{FB,\mu}$, $A_{LR,h}$ ,$R^{(5)}$, two percent for  $R_{b,\mu}$
and five percent for $A_{\tau}$. Results are shown in Fig.~30 for both models,
the different curves corresponding to the various observables, and the shaded
area to the combined overall mass bounds. In model (I) the resulting bounds
on $M_{V,A} $ are located in the 2 TeV range, and are rather strongly
correlated; the only hadronic observable which contributes appreciably is
$A_{LR,h}$  and allows to improve the pure leptonic result by about 200 GeV.
In model (II) the effect of releasing the validity of the Weinberg sum rule is
roughly that of increasing the bounds on $(M_V,M_A)$ from the
2 TeV to the 4 TeV region. Compared to the results obtained in
\cite{strong}, an improvement by a factor 6--8 as compared to the
LEP2 case. The explored mass range of $M_V/M_A$
should be able to give a definite hint of the existence of
Technicolour-like resonances or of any other strongly
coupled vector boson.

\newpage

\def\MPL #1 #2 #3 {Mod.~Phys.~Lett.~{\bf#1},\ #2 (#3)}
\def\NPB #1 #2 #3 {Nucl.~Phys.~{\bf#1},\ #2 (#3)}
\def\PLB #1 #2 #3 {Phys.~Lett.~{\bf#1},\ #2 (#3)}
\def\PR #1 #2 #3 {Phys.~Rep.~{\bf#1},\ #2 (#3)}
\def\PRD #1 #2 #3 {Phys.~Rev.~{\bf#1},\ #2 (#3)}
\def\PRL #1 #2 #3 {Phys.~Rev.~Lett.~{\bf#1},\ #2 (#3)}
\def\RMP #1 #2 #3 {Rev.~Mod.~Phys.~{\bf#1},\ #2 (#3)}
\def\ZP #1 #2 #3 {Z.~Phys.~{\bf#1},\ #2 (#3)}
\def\IJMP #1 #2 #3 {Int.~J.~Mod.~Phys.~{\bf#1},\ #2 (#3)}

\bibliographystyle{unsrt}

\begin{thebibliography}{99}
\bibitem{Sexp} We use the values for $\epsilon_3$ given by
G. Altarelli, CERN-TH-7319/94.
\end{thebibliography}

\end{document}